\documentclass[aps,superscriptaddress,twocolumn,twoside,floatfix,pra,nofootinbib,a4paper,notitlepage]{revtex4-1}
\usepackage{times}
\usepackage{epsfig}
\usepackage{amsfonts}
\usepackage{amsmath}
\usepackage{amssymb}
\usepackage{amsthm}
\newtheorem{theorem}{Theorem}

\newtheorem{lemma}{Lemma}
\newtheorem{proposition}{Proposition}
\usepackage{color}
\usepackage{multirow}
\usepackage{latexsym}
\usepackage{mathrsfs}
\usepackage{natbib}
\usepackage{verbatim}
\usepackage[utf8]{inputenc}
\usepackage[T1]{fontenc}
\usepackage{float}
\usepackage{dsfont}

\usepackage{array}[=2016-10-06]
\usepackage{booktabs}

\usepackage{mathtools}

\usepackage{graphicx}
\usepackage{xcolor}

\usepackage{wasysym}
\usepackage{bm}

\usepackage{tikz}
\usetikzlibrary{patterns,arrows.meta}

\newcommand{\ket}[1]{|#1\rangle}
\newcommand{\bra}[1]{\langle#1|}

\def\one{\leavevmode\hbox{\small1\normalsize\kern-.33em1}}

\usepackage[colorlinks=true,linkcolor=blue,citecolor=magenta,urlcolor=blue]{hyperref}
\renewcommand{\doibase}[1]{\if\space#1\expandafter\doibase\else https://doi.org/#1\fi}

\DeclareMathOperator{\Tr}{tr}
\newcommand{\Vopt}{V_{\mathrm{opt}}}
\newcommand{\Vmax}{V_{\max Q}}
\newcommand{\gap}[2]{#1\!\times\!10^{#2}}
\newcommand{\I}{\mathds{1}}

\begin{document}

\title{Facets and detection efficiency of prepare-and-measure correlations}

\author{Emmanuel Zambrini Cruzeiro}
\affiliation{Quantum Physics of Information (QPI) Group, Instituto de Telecomunicações, Lisbon, Portugal}
\affiliation{Quantum Information and Quantum Optics (QIQO) Laboratory, Instituto Superior Técnico, Lisbon, Portugal}

\begin{abstract}
    We present complete and partial lists of facet inequalities for prepare-and-measure scenarios with bit or trit communication and binary or ternary measurement outcomes. We compute classical and signalling bounds and obtain lower and upper bounds on the quantum values for qubits and qutrits. We also study their robustness to white noise and the detection efficiency required to violate the inequalities. We determine the exact maximum number of essential measurement inputs in a classical facet for any fixed number of preparations, message dimension, and number of outcomes, and construct facet families attaining this limit. For four preparations and four binary measurements, we obtain a qutrit critical detection efficiency of approximately $0.2779$ for violating a classical-bit facet. Under the same uniform-loss model, classical trits cannot violate any classical-bit inequality at efficiencies at or below $1/3$. Among the facets studied with four preparations and three measurements, allowing a third outcome lowers the qutrit detection threshold against a classical trit from approximately $0.76980036$ to $0.76822610$. No improvement in visibility larger than $10^{-5}$ is found. We also analytically construct a facet with ten preparations and three binary measurements achieving a qubit detection efficiency of approximately $0.6374$. This improves on the previously reported achievable detection threshold while substantially reducing the number of preparations and measurement settings.
\end{abstract}

	\maketitle

\section{Introduction}

Prepare-and-measure (PM) experiments constitute the elementary one-way communication scenario: a sender receives an input \(x\), encodes it into a physical system, and a receiver chooses a measurement \(y\) producing an outcome \(b\). The observed data are the conditional probabilities
\begin{equation}
    p(b|x,y),
\end{equation}
with \(x\in[X]\), \(y\in[Y]\), and \(b\in[B]\), where $[N]\coloneqq\{1,\ldots,N\}$. Without a restriction on the communicated system this scenario is trivial: if Alice can transmit \(x\) itself, Bob can reproduce any conditional distribution \(p(b|x,y)\). Nontrivial PM correlations therefore arise only once the communicated system is constrained.

The most widely studied constraint is a bound on the dimension of the communicated system. We denote by \(\mathcal C_M(X,Y,B)\) the set of classical PM correlations achievable when Alice sends an \(M\)-valued classical message, and by \(\mathcal Q_d(X,Y,B)\) the corresponding quantum set obtained when Alice sends a \(d\)-dimensional quantum system. Dimension-bounded PM scenarios provide the natural framework for dimension witnesses \cite{Hendrych2012,Ahrens2012,Brunner2013,Bowles2014,Cong2017,Aguilar2018}, semi-device-independent randomness generation \cite{Li2011}, and semi-device-independent quantum key distribution \cite{Pawlowski2011,Woodhead2015}. They have also been used to certify measurement properties beyond the observed statistics, including genuinely non-projective measurements and rank constraints on measurement effects \cite{Mironowicz2019,Tavakoli2020NonProjective,Steinberg2021,Navascues2023PM}. Other physically motivated restrictions on the communicated system have also been studied, including entropy constraints on the message \cite{Chaves2015,Tavakoli2020b,Tavakoli2022}, bounds on state overlaps, and energy constraints on the source \cite{VanHimbeeck2017}. For a recent overview of different restrictions in PM scenarios, see Ref.~\cite{pauwels2025information}.

The central object of this work is the classical bounded-message polytope \(\mathcal C_M(X,Y,B)\). Its facets are the tight linear constraints obeyed by all classical PM strategies with an \(M\)-valued message. Equivalently, they are tight classical dimension witnesses against which quantum \(d\)-dimensional systems may be tested. In this sense, facet inequalities in PM scenarios play a role analogous to tight Bell inequalities in nonlocality: they expose the irredundant linear structure of the relevant classical set and support the identification of minimal scenarios in which quantum advantages can occur. They are therefore natural candidates for robust quantum violations, for experimental dimension certification, and for identifying operational regimes in which bounded quantum communication outperforms bounded classical communication.

Despite this fundamental role, few complete classifications of PM facets are currently known. Ref.~\cite{Gallego2010} identified the single nontrivial facet of the simplest nontrivial PM polytope, namely \(\mathcal C_2(3,2,2)\), corresponding to three preparations, two binary measurements, and a binary classical message. This inequality, denoted \(I_3\), belongs to a family \(I_N\) of dimension witnesses, which was shown to define facets in several small cases. The inequality \(I_N\) has since served as a standard benchmark for experimental demonstrations of semi-device-independent dimension certification \cite{Hendrych2012,Ahrens2012}. Later, Ref.~\cite{Poderini2020} provided complete classifications for \(\mathcal C_2(3,3,2)\) and \(\mathcal C_2(4,2,2)\), finding only two and one new non-lifted nontrivial facet classes, respectively. This scarcity contrasts sharply with the Bell scenario, where extensive catalogues of facet classes are known for several small cases \cite{Avis2007,Cope2019,cruzeiro2019complete,Jesus2022,staufenbiel2026bell}. It leaves open a basic general question: what are the tight constraints governing bounded one-way classical communication?

In this work we systematically characterize classical correlation polytopes in dimension-bounded PM scenarios. We enumerate all nontrivial facet classes of \(\mathcal C_M(X,Y,B)\) for various scenarios and also provide extensive symmetry-reduced partial catalogues. Our facet catalogues cover $M,B\in\{2,3\}$, and we restrict the numerical quantum optimization to qubits and qutrits, $d\in\{2,3\}$. Strategies using a classical quart (a four-valued message) also determine quantum score and detection-efficiency bounds for four preparations, without further quantum optimization. The analytical results hold for the parameter ranges stated in each theorem.
The facet representatives, symmetry data, and numerical bounds will be deposited in an accompanying public repository. Completeness is certified when an exhaustive facet-enumeration algorithm runs to completion. Without such verification, including for catalogues obtained through adjacency sampling~\cite{staufenbiel2026bell}, we distinguish conjecturally complete lists from partial ones according to the available computational evidence.

A structural question underlying these catalogues is how many of Bob's measurements can be essential in a tight classical dimension witness. We determine the exact input-saturation number for every nontrivial message dimension and every number of outcomes. Beyond this cutoff, every facet is obtained by input lifting from a scenario with fewer measurements. Matching facet constructions show that the cutoff cannot be reduced, and that it is independent of the number of outcomes. This result gives an exact stopping point for the appearance of new facets as Bob's number of inputs increases, at fixed preparation number, message dimension, and outcome number.

We then analyze the operational relevance of the resulting facets. For each inequality we compute lower bounds on quantum violations using see-saw optimization and upper bounds using the Navascués--Vértesi hierarchy. We also evaluate the corresponding noise robustness and critical detection efficiency. The full conditional-probability polytope is recovered for \(M\geq\min\{X,B^Y\}\); in particular, \(M\geq X\) allows Alice to communicate her input exactly. Saturating an individual witness does not imply that every unrestricted behavior can be reproduced. This analysis reveals several minimal scenarios in which \(d\)-dimensional quantum systems outperform classical messages of the same dimension. These results provide a systematic catalogue of tight PM witnesses for experimentally relevant quantum communication scenarios.

We then focus on the qubit case. For nontrivial binary projective measurements, we find no analyzed facet with a white-noise visibility below the standard random access code (RAC)/CHSH value $1/\sqrt{2}$. Exclusion of such an improvement is certified in those scenarios for which both a complete facet list and sufficiently tight quantum upper bounds are available. In terms of detection efficiency, improvements are possible in small scenarios: the standard $I_4$ witness already gives $2/3$ with four preparations and three measurements~\cite{Gallego2010,Hendrych2012}, compared with $1/\sqrt{2}$ for the RAC in $(4,2,2,2)$~\cite{Li2015}. Here and below, a scenario is denoted by $(X,M,Y,B)$. Guided by the preparation and measurement structures of smaller examples, we analytically construct a facet in $(10,2,3,2)$ attaining $\eta_{\mathrm{crit}}=\eta_{\mathrm{crit}}^G\simeq0.6373835250$, with an exact optimum over general dichotomic qubit measurements. This improves the lower bound on the Grothendieck-like constant $K_D$ defined in Ref.~\cite{divianszky2023certification}. The construction of Divi\'anszky \textit{et al.}~\cite{divianszky2023certification} uses 70 preparations and 70 measurement settings, and reports an achievable detection threshold of approximately $0.63764$. Our construction improves on this reported threshold using only ten preparations and three measurements.

\section{Convex polytopes and facet enumeration}

\subsection{Basics of convex polytopes}

In this section, we briefly recall basic definitions pertaining to convex polytopes. A convex polytope can be described in multiple ways \cite{Ziegler1993}. Two important descriptions are
\begin{enumerate}
    \item $\mathcal{V}$-description: the \textit{convex hull} $\mathcal{C}$  of a set of points ${x_1,\dots,x_N}$, where $x_j\in\mathbb{R}^D$, $j=1,\dots,N$,
    \begin{equation}
    \mathcal{C} = \{\sum_j \lambda_j x_j | \lambda_j \geq 0, \sum_j \lambda_j = 1\}
    \end{equation}
    \item $\mathcal{H}$-description: a bounded intersection of finitely many half-spaces,
    \begin{equation}
    \{x \in \mathbb{R}^D: Ax \geq b\},\quad A\in\mathbb{R}^{m\times D},\quad b\in\mathbb{R}^m
    \end{equation}
\end{enumerate}

The main theorem for polytopes \cite{Ziegler1993} states that a set is the convex hull of finitely many points if and only if it is a bounded intersection of finitely many half-spaces.

For a valid supporting inequality $\alpha\cdot x\leq\beta$ of a convex polytope $P$, the associated face is
\begin{equation}
 \mathcal F=\{x\in P:\alpha\cdot x=\beta\},
 \qquad \alpha\in\mathbb R^D,\quad\beta\in\mathbb R.
\end{equation}
A facet is a face of dimension $\dim P-1$, and its supporting inequality is called facet-defining.

There are infinitely many $\mathcal H$-descriptions of a convex polytope. A minimal description consists of its affine-hull equations together with its facet-defining inequalities.

In both Bell and PM polytopes, the $\mathcal{V}$-description is easier to generate than the $\mathcal{H}$-description, as it amounts to generating all deterministic mappings for a given scenario. The number of vertices and the dimension of the polytope increase rapidly with the number of preparations and settings, so generating all the vertices eventually becomes hard when we increase the number of settings and outcomes. Obtaining the $\mathcal{H}$-description is more difficult: guessed or constructed inequalities need not define facets. On the other hand, given the complete $\mathcal{V}$-description, a valid inequality is facet-defining precisely when its saturating vertices have affine rank $\dim P-1$. The problem of converting a complete set of vertices to the minimal $\mathcal{H}$-description is called the \textit{facet enumeration problem}. Several software tools are available to solve this problem, such as lrs \cite{Avis2000}, PORTA \cite{Christof1997}, PANDA \cite{Lorwald2015}.

\subsection{Facet enumeration via liftings and polytope slicing}
\label{subsec:enumeration}

As is well-known, direct facet enumeration is limited. For example, in bipartite Bell scenarios, and for binary outcomes, the full list of facet classes for four settings on each side was presented in \cite{oudot2019bipartite,cruzeiro2019complete}. But deriving this list from direct facet enumeration is computationally intractable. However, as shown in \cite{Jesus2022}, one can actually find the list of facets and show that it is complete in a reasonable time by \textit{slicing} the polytope.

The slicing concept is straightforward: one slices a polytope into smaller pieces using half-spaces to split the vertices into two, or more, sets. As an illustration, by choosing a hyperplane $A$ and a bound $\beta$, we may split the complete set of vertices for a given scenario into two sets $C_1$ and $C_2$ of indices, where vertices indexed by values in the first set satisfy $A x_i \leq \beta$, for all $i\in C_1$, while the others don't.

For each selected vertex subset, we enumerate the facets of its convex hull. This operation differs from intersecting the original polytope with a half-space, which can introduce new vertices at cut-edge crossings. Each resulting candidate is checked for validity against all original vertices and for facetness by the affine-rank test. By taking many hyperplanes, specifically liftings in the procedure in \cite{Jesus2022}, and setting different bounds, one can reduce the sizes of these auxiliary enumeration problems. Completeness is certified only when the subsequent exhaustive enumeration runs to completion.

PANDA enables speedups in facet enumeration when supplied with known facets, for example liftings~\cite{pironio2005lifting} or facets obtained by enumeration of slices of the original polytope. For PM scenarios at fixed message dimension $M\ge2$, preparation and measurement-input liftings assign zero coefficients to the added coordinates and preserve facets. Outcome liftings copy the coefficients of a split outcome and preserve non-positivity facets. For completeness, proofs of these three lifting properties are given in Appendix~\ref{app:liftings}; related signaling-polytope liftings appear in Ref.~\cite{Doolittle2021Lifting}.

We use a combination of facet enumeration methods. When attempting a complete classification, we use the following methods in order, depending on the complexity of the problem:
\begin{enumerate}
    \item Exhaustive facet enumeration using lifted facets as priors.
    \item Exhaustive facet enumeration using lifted facets and facets obtained by enumeration of slices of the original polytope as priors.
\end{enumerate}
When complete enumeration is impractical, we combine adjacency sampling~\cite{staufenbiel2026bell} and slicing to obtain partial lists of facets.

\section{Dimension-bounded prepare-and-measure scenarios}

\subsection{Classical correlations and probability representation}

In the classical setting, Alice sends a classical message $m\in [M]$ to Bob. Bob outputs a result $b\in [B]$ based on his input $y\in [Y]$ and the message $m$. With unrestricted shared randomness, the resulting correlations are
\begin{equation}
p(b|x,y) = \sum_\lambda p(\lambda) \sum_{m=1}^M p^A(m|x,\lambda)p^B(b|m,y,\lambda).
\end{equation}

To describe the set of probabilities in a dimension-bounded scenario, we must choose a representation to work with. The normalized conditional probability distributions
$\{p(b|x,y)\}_{b,x,y}$ belong to an affine space of dimension
$(B-1)XY$.

The most direct representation, which we call full-probability
(or full-$P$) notation, retains all conditional probabilities
\begin{equation}
    \{p(b|x,y)\}_{b\in[B],\,x\in[X],\,y\in[Y]}.
\end{equation}
This representation is redundant because of normalization.

Using normalization, we can eliminate one outcome probability for each pair of preparation and measurement inputs, leaving $(B-1)XY$ independent variables. We conventionally omit the last outcome and retain
\begin{equation}
    \{p(b|x,y)\}_{b\in[B-1],\,x\in[X],\,y\in[Y]}.
\end{equation}

For binary outcomes, we also use correlator notation, relabeling outcomes $1$ and $2$ as $+1$ and $-1$, respectively. The corresponding expectation values are
\begin{equation}
    E_{x,y} = \sum_{b = \pm 1} b p(b|x,y) = p(+1|x,y) - p(-1|x,y).
\end{equation}

The full probability notation is useful because any symmetry (relabeling of inputs and outcomes) consists only in a permutation of the labels.

Finally, correlator notation is particularly useful for binary-outcome scenarios, since visibility calculations often take a simpler form. For qubit preparations and nontrivial binary projective measurements, it also makes the connection to Grothendieck-type optimization explicit~\cite{divianszky2023certification}.

Since the classical correlation set is the convex hull of deterministic
strategies, its vertices may be represented by deterministic encoding
and decoding functions
\begin{equation}
    f:[X]\rightarrow[M],
    \qquad
    g:[M]\times[Y]\rightarrow[B].
\end{equation}
Different pairs $(f,g)$ can give the same vertex. The distinct vertices are exactly the deterministic outcome tables with at most $M$ distinct response strings across Alice's preparations. For $M\geq2$ and $B\geq2$, their affine dimension is $D=XY(B-1)$: the all-reference-output table and the tables obtained by changing one entry span the normalized probability space and use at most two messages. For $M=1$, all preparation rows coincide.

The set of correlations attainable with a classical message $m\in[M]$ is a convex polytope, denoted by $\mathcal{C}_M$. The facets of such an object are linear inequalities
\begin{equation}
I_{\boldsymbol{c}}[p] := \sum_{b,x,y}c_{bxy}p(b|x,y) \leq C_M(\boldsymbol{c}),
\end{equation}
where
\begin{equation}
\boldsymbol{c} = \{c_{bxy}\}_{b\in[B],\,x\in[X],\,y\in[Y]}
\end{equation}
is the corresponding array of real coefficients, and
$C_M(\boldsymbol{c})$ is the classical bound for messages of
cardinality $M$ in full-probability notation. It is computed in the
following way:
\begin{align}
C_M(\boldsymbol{c}) &= \max_{p\in\mathcal{C}_M} \sum_{b,x,y}c_{bxy}p(b|x,y), \\
&=
\max_{p^A,p^B\in\{0,1\}} \sum_{b,x,y}c_{bxy} \sum_m p^A(m|x)p^B(b|m,y),\\
&= \max_{f,g} \sum_{b,x,y}c_{bxy} \sum_m\delta_{m,f(x)}\delta_{b,g(m,y)}.
\end{align}
where the maximum is taken over all deterministic strategies, i.e. over $p^A(m|x,\lambda)\in\{0,1\}$ and $p^B(b|m,y,\lambda)\in\{0,1\}$.

For binary outcomes, with outcome $1$ corresponding to $+1$, define
\begin{equation}
 \widetilde c_{xy}=\frac{c_{1xy}-c_{2xy}}2,
 \qquad a=\frac12\sum_{x,y}(c_{1xy}+c_{2xy}).
\end{equation}
Then $I_{\boldsymbol c}[p]=a+\sum_{x,y}\widetilde c_{xy}E_{xy}$, so all bounds and white-noise contributions shift by the same constant $a$.

If we change to correlator notation, the functional becomes
\begin{equation}
I_{\widetilde{\boldsymbol{c}}}[E] := \sum_{x,y}\tilde{c}_{xy}E_{xy},
\end{equation}
where
\begin{equation}
\widetilde{\boldsymbol{c}} = \{\tilde{c}_{xy}\}_{x\in[X],\,y\in[Y]}
\end{equation}
is the corresponding array of real coefficients. The classical bound is then
\begin{equation}
\widetilde{C}_M(\widetilde{\boldsymbol{c}}) = \max_{E\in\widetilde{\mathcal{C}}_M} I_{\widetilde{\boldsymbol{c}}}[E] = \max_{E\in\widetilde{\mathcal{C}}_M}\sum_{x,y}\widetilde{c}_{xy}E_{xy},
\end{equation}
where $\widetilde{\mathcal{C}}_M$ is the classical polytope in correlator space.

\subsection{Symmetries of the scenario}

The symmetry group of a prepare-and-measure scenario partitions the
facets into equivalence classes. Alice's inputs and Bob's inputs may be
permuted independently, while Bob's outcomes may be permuted
independently for each value of $y$. We therefore report one
representative from each facet-equivalence class. The relabeling group is $S_X\times((S_B)^Y\rtimes S_Y)$. In binary correlator notation, this allows row and column permutations and independent column sign flips; independent row sign flips and exchange of Alice and Bob are not generally PM symmetries. Inequalities are also identified modulo normalization identities and positive rescaling, which must be fixed when comparing coefficient representatives.

In full-probability notation, these relabelings act as permutations of
the probability coordinates and, consequently, as permutations of the
inequality coefficients.

\subsection{Quantum correlations}

A finite-dimensional quantum state is a density matrix $\rho \in \mathcal{D}(\mathbb{C}^d)$, where
\begin{equation}
    \mathcal{D}(\mathbb{C}^d) = \{ \rho\in\mathbb{C}^{d\times d}: \rho\succeq 0, \text{Tr}[\rho]=1 \}.
\end{equation}

A general finite-dimensional $K$-outcome quantum measurement, also called a positive operator-valued measure (POVM), is a set of matrices $\{M_k\}_{k\in[K]}$ satisfying
\begin{equation}
    M_k\in\mathbb{C}^{d\times d},\qquad M_k \succeq 0,\qquad \sum_k M_k = \mathds{1}_d.
\end{equation}

A more restrictive type of measurement is the projection-valued measure (PVM), i.e., a POVM $\{M_k\}_{k\in[K]}$ satisfying $M_kM_{k'}=\delta_{k,k'}M_k$. Unless a rank restriction is stated, zero projectors and all projector ranks are allowed. In particular, a three-outcome qubit PVM has at least one zero projector; this class is distinct from the binary rank-one class denoted by $G$ below.

In the quantum PM scenario, Alice receives an input $x\in[X]$, prepares a state $\rho_x\in\mathcal{D}(\mathbb{C}^d)$, and sends it to Bob. Bob receives an input $y\in[Y]$ and performs the corresponding POVM $\{M_{b|y}\}_{b\in[B]}$. Whenever an average score requires input distributions, we assume that Alice's and Bob's inputs are independently and uniformly distributed. With arbitrary shared randomness, the observed correlations are
\begin{equation}
p(b|x,y) = \sum_\lambda p(\lambda)\text{Tr} [\rho_x^{(\lambda)}M^{(\lambda)}_{b|y}],
\end{equation}
where $\lambda$ is a shared classical variable, independent of $x$ and $y$, with distribution $p(\lambda)$.

Since the objective is linear in each preparation, its maximum can
always be attained with pure states, i.e., rank-one density matrices. Similarly, it is sufficient to optimize over extremal POVMs~\cite{divianszky2023certification}. In binary-outcome scenarios, every extremal POVM is a PVM, and hence the optimization may be restricted to projective measurements, including the deterministic measurements $\{\mathds{1}_d,0\}$ and $\{0,\mathds{1}_d\}$.

We define the quantum $d$-dimensional bound of a prepare-and-measure inequality,
\begin{equation}
Q_d(\boldsymbol{c}) = \max_{\substack{ \rho_x\in\mathcal{D}(\mathbb{C}^d)\ \forall x\\ \{M_{b|y}\}_{b\in[B]}\ \mathrm{POVM}\ \forall y}} \sum_{b,x,y} c_{bxy}\operatorname{Tr}\!\left[M_{b|y}\rho_x\right].
\end{equation}
Since the functional is linear, shared randomness cannot increase its maximum and is therefore omitted from the optimization defining $Q_d$.

For binary-outcome inequalities, the optimum can be attained with pure states and PVMs, although the optimal projectors need not have rank one. More generally, for scenarios with $B=d$, we define the value restricted to pure preparations and rank-one PVMs as
\begin{align}
q_d(\boldsymbol{c})
={}&
\max_{\{\ket{\psi_x}\},\,\{\ket{\phi_{b|y}}\}}
\sum_{b,x,y}
c_{bxy}
\left|\langle\phi_{b|y}|\psi_x\rangle\right|^2
\\
\text{subject to}\quad
&\ket{\psi_x}\in\mathbb{C}^d,\qquad
\langle\psi_x|\psi_x\rangle=1,
&&\forall x,
\\
&\ket{\phi_{b|y}}\in\mathbb{C}^d,
&&\forall b,y,
\\
&\langle\phi_{b|y}|\phi_{b'|y}\rangle
=\delta_{bb'},
&&\forall b,b',y,
\\
&\sum_{b=1}^{d}
\ket{\phi_{b|y}}\!\bra{\phi_{b|y}}
=
\mathds{1}_d,
&&\forall y,
\end{align}
where $M_{b|y}=\ket{\phi_{b|y}}\!\bra{\phi_{b|y}}$.

For $B=d$,
\begin{equation}
q_d(\boldsymbol{c}) \leq Q_d(\boldsymbol{c}).
\end{equation}

For binary outcomes, we will sometimes use the correlator notation. Since passing between full-probability and correlator notation generally entails an affine transformation of the functional, the corresponding numerical bounds depend on the chosen representation and normalization. For the most general binary-outcome quantum bound in correlator notation, we will use

\begin{align}
\widetilde{Q}_d(\widetilde{\boldsymbol c})
={}&
\max_{\{\rho_x\},\,\{A_y\}}
\sum_{x,y}
\tilde{c}_{xy}\operatorname{Tr}\!\left(\rho_xA_y\right)
\\
\text{subject to}\quad
&\rho_x\in\mathcal{D}(\mathbb{C}^d),
&&\forall x,
\\
&A_y=A_y^\dagger,
&&\forall y,
\\
&-\mathds{1}_d\preceq A_y\preceq\mathds{1}_d,
&&\forall y.
\end{align}

For qubit preparations, the rank-one PVM bound for the correlator notation becomes particularly elegant:
\begin{align}
\widetilde{q}_2(\widetilde{\boldsymbol c}) ={}& \max_{\{\vec{a}_x\},\,\{\vec{b}_y\}} \sum_{x,y} \tilde{c}_{xy}\,\vec{a}_x\cdot\vec{b}_y \\
\text{subject to}\quad
&\vec{a}_x\in\mathbb{R}^3,\quad \|\vec{a}_x\|=1,
&&\forall x, \\
&\vec{b}_y\in\mathbb{R}^3,\quad \|\vec{b}_y\|=1,
&&\forall y.
\end{align}
where $\vec{a}_x, \vec{b}_y$ are Bloch vectors characterizing the states and measurements, respectively.

Then
\begin{equation}
\tilde{q}_2(\widetilde{\boldsymbol c}) \leq \tilde{Q}_2(\widetilde{\boldsymbol c}).
\end{equation}
The restricted qubit value excludes deterministic measurement settings; any improvement in the unrestricted binary optimum can be attained with PVMs, without requiring genuinely nonprojective measurements.

For binary measurements in dimension $d$, we separately define
\begin{align}
 \widetilde q_{d,r}(\widetilde{\boldsymbol c})
 &=\max_{\{\rho_x\},\{P_y\}}
 \sum_{x,y}\widetilde c_{xy}\Tr[\rho_x(2P_y-\mathds{1}_d)],
 \\
 &\rho_x\in\mathcal D(\mathbb C^d),\quad
 P_y=P_y^\dagger=P_y^2,\quad\operatorname{rank}P_y=r.
\end{align}
Thus the $+1$ effect has rank $r$ and the $-1$ effect rank $d-r$. We use $r=1$ for the rank-restricted binary values, with $\widetilde q_{2,1}=\widetilde q_2$. This binary restriction is distinct from the $B=d$ rank-one PVM quantity $q_d$. Under an outcome relabeling, the two effect ranks must also be exchanged.

To obtain feasible lower bounds on the quantum values, we use an iterative semidefinite-programming see-saw optimization over states and measurements. For upper bounds, we rely on the Navascués--Vértesi (NV) hierarchy~\cite{Navascues2015,Navascues2015b}. Bounds are compared within the same measurement class: rank-restricted optimizations must impose the stated ranks, and a PVM relaxation for $B>2$ is not automatically an upper bound for arbitrary $d$-dimensional POVMs.
\subsection{Signalling bound}

The signalling, or algebraic, bound is the maximum value of the functional over arbitrary conditional probability distributions. Equivalently, it is the value obtained when Bob may choose his output as a function of both inputs \(x\) and \(y\), corresponding to unrestricted communication from Alice to Bob. In full probability notation, the signalling bound is
\begin{equation}
S(\{\boldsymbol{c}\})
=
\sum_{x=1}^{X}\sum_{y=1}^{Y}
\max_{b\in[B]} c_{bxy}.
\end{equation}

For a correlator functional
\begin{equation}
I_{\boldsymbol{\tilde{c}}}[E] = \sum_{x,y}\tilde{c}_{xy}E_{xy},
\end{equation}
the signalling bound is
\begin{equation}
\tilde{S}(\widetilde{\boldsymbol c}) = \sum_{x=1}^{X}\sum_{y=1}^{Y}
|\tilde{c}_{xy}|.
\end{equation}

\subsection{Resistance to white noise}

Let
\begin{equation}
I[p] = \sum_{b,x,y} c_{bxy}p(b|x,y)
\end{equation}
be a prepare-and-measure functional, and denote by $Q$ its value for a
given noiseless quantum realization. We model white noise on the prepared states by the depolarizing map~\cite{divianszky2023certification,Li2015}
\begin{equation}
\rho_x(V)
=
V\rho_x
+
(1-V)\frac{\mathds{1}_d}{d}.
\end{equation}
The value of the functional becomes
\begin{equation}
I(V)
=
VQ+(1-V)I_{\mathrm{wn}},
\end{equation}
where
\begin{equation}
I_{\mathrm{wn}} = \sum_{b,x,y} c_{bxy} \operatorname{Tr}\!\left( \frac{\mathds{1}_d}{d}M_{b|y} \right)
\end{equation}
is the value obtained from maximally mixed preparations. If $C$ is the corresponding classical bound, the critical visibility of this realization is therefore
\begin{equation}
V_{\mathrm{crit}} = \frac{C-I_{\mathrm{wn}}}{Q-I_{\mathrm{wn}}},
\end{equation}
provided that $Q>C\geq I_{\mathrm{wn}}$. The tolerated white-noise fraction is $1-V_{\mathrm{crit}}$. Since the maximally mixed-state behavior is independent of $x$, it is classically realizable with one message and satisfies $I_{\mathrm{wn}}\leq C$.

For projective measurements with fixed outcome ranks, the white-noise contribution is constant, so maximizing the quantum violation also minimizes the critical visibility. For the binary class $\widetilde q_{d,r}$,
\begin{equation}
 I_{\mathrm{wn}}=a+\frac{2r-d}{d}\sum_{x,y}\widetilde c_{xy},
\end{equation}
where $a$ is the affine offset in the full-probability witness. For the shifted correlator functional, $a=0$. Thus rank one does not imply a zero white-noise contribution outside the qubit case. For unrestricted POVMs, the white-noise contribution can vary with the measurements, so optimal visibility requires optimizing the ratio above, or the noisy score at fixed $V$. For binary outcomes, projectors suffice at each fixed $V$, but their different rank patterns must be compared. A complete facet list remains sufficient when optimizing realizations: every nonclassical noisy behavior violates a facet. We therefore distinguish two optimizations. For a realization $R$, let $S_R$ be its noiseless score and $W_R$ its white-noise score, and write $v_R=(C_M-W_R)/(S_R-W_R)$. Within a specified dimension and measurement class, define
\begin{align}
 \Vopt&=\min_{R:S_R>C_M}v_R,\label{eq:Vopt}\\
 \Vmax&=\min_{R:S_R=Q_d}v_R,\qquad Q_d>C_M.\label{eq:Vmax}
\end{align}
The second quantity selects the smallest visibility among realizations attaining the exact noiseless maximum. Equivalently, it selects the largest $W_R$ among those realizations. A near-maximal score alone does not certify membership in this set. A feasible realization supplies an upper bound on $\Vopt$, but supplies one on $\Vmax$ only if it is also an exact noiseless maximizer. We report separate intervals when the numerical precision makes the distinction relevant; a shared $V$ column denotes agreement of the two scenario minima within the stated tolerance, not equality for every facet.

For qubit correlator inequalities evaluated with nontrivial binary projective measurements, the observables are traceless and hence $I_{\mathrm{wn}}=0$. The superscript $G$ will denote this Grothendieck-related qubit measurement class. Optimizing over this class gives
\begin{equation}
V_{\mathrm{crit}}^G(\{\widetilde{c}_{xy}\}) = \frac{\widetilde{C}_2(\{\widetilde{c}_{xy}\})}{\widetilde{q}_2(\{\widetilde{c}_{xy}\})}.
\end{equation}

Let $G=(G_{xy})\in\mathbb{R}^{X\times Y}$ be the real coefficient matrix of a binary-outcome correlator functional,
\begin{equation}
I_G
=
\sum_{x=1}^{X}\sum_{y=1}^{Y}
G_{xy}E_{xy}.
\end{equation}
We define
\begin{equation}
\widetilde{q}_2(G)
=
\max_{\substack{
\vec{a}_x,\vec{b}_y\in\mathbb{R}^3\\
\|\vec{a}_x\|=\|\vec{b}_y\|=1
}}
\sum_{x=1}^{X}\sum_{y=1}^{Y}
G_{xy}\,
\vec{a}_x\cdot\vec{b}_y,
\end{equation}
and
\begin{equation}
\widetilde{C}_2(G)
=
\max_{E\in\widetilde{\mathcal{C}}_2}
\sum_{x=1}^{X}\sum_{y=1}^{Y}
G_{xy}E_{xy}.
\end{equation}
The prepare-and-measure Grothendieck ratio is then
\begin{equation}
K_{\mathrm{PM}}
=
\sup_{\substack{
X,Y\in\mathbb{N}\\
G\in\mathbb{R}^{X\times Y}\\
G\neq 0
}}
\frac{\widetilde{q}_2(G)}{\widetilde{C}_2(G)}.
\end{equation}

Following \cite{divianszky2023certification},
\begin{equation}
    K_\text{PM} = K_G(3).
\end{equation}
Hence, the reciprocal of the critical visibility of any binary-outcome correlator witness with pure states and nontrivial rank-one projective measurements provides a lower bound
\begin{equation}
\frac{1}{V^G_{\mathrm{crit}}} \leq K_G(3),
\end{equation}
or, equivalently, an upper bound on the optimal critical visibility $1/K_G(3)$.

Combining the relation $K_{\mathrm{PM}}=K_G(3)$ with the currently known bounds \cite{designolle2023improved,designolle2026better} on $K_G(3)$ yields
\begin{equation}
    0.6875 \lesssim 1/K_G(3) \lesssim 0.69604,
\end{equation}
where the decimal endpoints represent rounded bounds. The refined lower bound $K_G(3)\gtrsim1.43670$ in Ref.~\cite{designolle2026better} is obtained from a $97\times97$ Bell matrix. Applying the matrix-and-negative stacking construction of Ref.~\cite{divianszky2023certification} gives a $194\times97$ PM matrix; the visibility near $0.69604$ should therefore be distinguished from the smaller $48\times24$ construction.

Recent bounds on the unrestricted real constant $K_G=\sup_nK_G(n)$, including $1.773\leq K_G\leq1.7799$~\cite{heilman2026primaldual}, and bounds on the complex Grothendieck constant~\cite{heilman2026complex} concern different constants. Conditional finite-order proposals based on conjectured binary optima~\cite{bene2026heuristic} do not replace the unconditional order-three bounds used here. These results therefore leave the qubit visibility benchmark unchanged. The order of a Grothendieck constant is a real vector dimension and should not be identified automatically with the communicated Hilbert-space dimension.

In this work, we compute visibilities for the facets we obtain to see how close they are to the best currently known bounds associated to $K_G(3)$. This is relevant for experimental implementations of prepare-and-measure scenarios. An interesting question for pure qubit preparations and nontrivial binary projective measurements is how small a scenario can achieve $V<1/\sqrt2$; Ref.~\cite{divianszky2023certification} gives a construction in $(48,2,24,2)$. For all the $(X,2,Y,2)$ facets we have analyzed, we have found no example with a visibility lower than $1/\sqrt{2}$. Our search does not settle the minimal scenario size. We also investigate more general dimension-dependent quantum-to-classical ratios.

We define more general quantities comparing $d$-dimensional quantum systems with a classical bit,
\begin{equation}
K_\text{PM}^d =
\sup_{\{\widetilde{c}_{xy}\}}
\frac{
\widetilde{Q}_d(\{\widetilde{c}_{xy}\})
}{
\widetilde{C}_2(\{\widetilde{c}_{xy}\})
}.
\end{equation}
We use this ratio as a dimension-dependent measure of quantum advantage. For a fixed realization whose white-noise score vanishes, the critical visibility against a classical bit is $\widetilde C_2/Q$.

Similarly, for the rank-restricted binary class, we define
\begin{equation}
K_\text{PM}^{d,r} =
\sup_{\{\widetilde{c}_{xy}\}}
\frac{\widetilde q_{d,r}(\{\widetilde{c}_{xy}\})}
{\widetilde C_2(\{\widetilde{c}_{xy}\})}.
\end{equation}
Both suprema range over arbitrary finite $X,Y$ and nonzero coefficient arrays. These are quantum-versus-bit comparisons, rather than equal-dimension comparisons for $d>2$.

A constant offset must be included when comparing representations. Under $I'=sI+t$, with $s>0$ and $t\in\mathbb R$ independent of the behavior, all scores and bounds transform in the same way. Consequently,
\begin{equation}
 \frac{C'-I'_{\mathrm{wn}}}{Q'-I'_{\mathrm{wn}}}
 =\frac{C-I_{\mathrm{wn}}}{Q-I_{\mathrm{wn}}},
\end{equation}
whereas a raw ratio $C'/Q'$ need not equal $C/Q$. The offset $t$ includes constants introduced by normalization or by converting probabilities to correlators. The same cancellation of the affine offset applies to the detection-efficiency ratio below.

\subsection{Detection efficiency}

A key figure of merit for prepare-and-measure facets is the minimum detection efficiency required to close the detection loophole. We assume uniform erasure with probability $1-\eta$, followed by a fixed output $b_y$ on a no-click event for measurement $y$. At the level of measurement effects, this is
\begin{equation}
 M^{(\eta)}_{b|y}=\eta M_{b|y}
 +(1-\eta)\delta_{b,b_y}\mathds{1}_d.
\end{equation}
Every trial is retained. If the ideal realization gives $Q$ and the no-click assignment gives $I_{\mathrm{nc}}$, the witness value under loss is
\begin{equation}
 I(\eta)=\eta Q+(1-\eta)I_{\mathrm{nc}}.
\end{equation}
The uniform-loss assumption specifies the quantum performance being evaluated; the classical bound applies to the full assigned-output data without postselection.

If $C$ is the classical bound, the critical detection efficiency of
this realization is
\begin{equation}
\eta_{\mathrm{crit}} = \frac{C-I_{\mathrm{nc}}}{Q-I_{\mathrm{nc}}},
\end{equation}
provided that $Q>C\geq I_{\mathrm{nc}}$. If $I_{\mathrm{nc}}=C$, the threshold is zero and every $\eta>0$ yields a violation.

For a binary-outcome correlator functional with coefficient array
$\widetilde{\boldsymbol{c}}$, suppose that a no-click event is assigned
the outcome $+1$. The no-click value is then
\begin{equation}
\Gamma(\widetilde{\boldsymbol{c}})
=
\sum_{x,y}\widetilde{c}_{xy}.
\end{equation}
Allowing a separate deterministic assignment for each measurement gives the optimal score
\begin{equation}
 \Gamma_\star(\widetilde{\boldsymbol c})
 =\sum_y\left|\sum_x\widetilde c_{xy}\right|.
\end{equation}
For the nontrivial binary qubit class, column relabelings preserve the measurement class and can make every column sum nonnegative, so that $\Gamma=\Gamma_\star$. Higher-dimensional rank restrictions need not be invariant under these relabelings; below we use $\Gamma_\star$ directly without changing the prescribed ranks.
Following Ref.~\cite{divianszky2023certification}, we define
\begin{equation}
K_D
=
\sup_{\widetilde{\boldsymbol{c}}}
\frac{
\widetilde{q}_2(\widetilde{\boldsymbol{c}})
-
\Gamma(\widetilde{\boldsymbol{c}})
}{
\widetilde{C}_2(\widetilde{\boldsymbol{c}})
-
\Gamma(\widetilde{\boldsymbol{c}})
}.
\end{equation}
The supremum is over arbitrary finite scenarios and coefficient arrays with $\widetilde C_2>\Gamma$, excluding undefined $0/0$ cases. The optimal threshold over violating witnesses is
\begin{equation}
\eta^G_\star
=
\inf_{\widetilde{\boldsymbol{c}}}
\frac{
\widetilde{C}_2(\widetilde{\boldsymbol{c}})
-
\Gamma(\widetilde{\boldsymbol{c}})
}{
\widetilde{q}_2(\widetilde{\boldsymbol{c}})
-
\Gamma(\widetilde{\boldsymbol{c}})
}
=
\frac{1}{K_D}.
\end{equation}
Here the infimum is restricted to $\widetilde q_2>\widetilde C_2>\Gamma$. For equal-dimension comparisons, define
\begin{equation}
\eta_{d,r}(\widetilde{\boldsymbol{c}})
=
\frac{
\widetilde{C}_d(\widetilde{\boldsymbol{c}})
-
\Gamma_\star(\widetilde{\boldsymbol{c}})
}{
\widetilde{q}_{d,r}(\widetilde{\boldsymbol{c}})
-
\Gamma_\star(\widetilde{\boldsymbol{c}})
},
\end{equation}
where $\widetilde q_{d,r}$ is the rank-restricted binary value defined above and the classical message dimension is now $d$. We use $r=1$ for these restricted values and retain the stated outcome/rank convention.

\begin{equation}
\eta_{d,Q}(\widetilde{\boldsymbol{c}})
=
\frac{
\widetilde{C}_d(\widetilde{\boldsymbol{c}})
-
\Gamma_\star(\widetilde{\boldsymbol{c}})
}{
\widetilde{Q}_d(\widetilde{\boldsymbol{c}})
-
\Gamma_\star(\widetilde{\boldsymbol{c}})
}.
\end{equation}
These expressions apply when the corresponding quantum value exceeds the classical bound. Since the optimal no-click score depends only on the witness, a maximally violating realization minimizes detection efficiency within any fixed measurement class, for any number of outcomes. A feasible quantum value yields an achieved threshold; matching quantum upper bounds certify its optimality. A proof valid for any number of outcomes is given in Appendix~\ref{app:ideal-loss}.

For nontrivial binary qubit projective measurements with deterministic assignment of no-click events, the bounds in Ref.~\cite{divianszky2023certification}, including its Eq.~(76), imply
\begin{equation}
 \frac12\leq\eta^G_\star<\frac{218298}{342353}
 \simeq0.6376401.
\end{equation}
The construction giving this upper bound uses 70 preparations and 70 measurement settings. The rational upper bound follows from an explicit achievable quantum value and an exact classical bound; it does not establish the minimum detection threshold of that particular witness. A small-scenario benchmark is the standard $I_4$ witness, with four preparations, three binary measurements, bit bound $5$, qubit value $6$, and optimal no-click score $3$~\cite{Gallego2010,Hendrych2012}. Thus, in the same loss model,
\begin{equation}
 \eta_{\mathrm{crit}}(I_4)=\frac{5-3}{6-3}=\frac23.
\end{equation}
Our $(10,2,3,2)$ witness reaches $\eta_{\mathrm{crit}}=\eta_{\mathrm{crit}}^G\simeq0.6373835250$ and gives the exact lower bound $K_D\geq2\sqrt3/3+\sqrt2-1\simeq1.5689141007$, derived in Sec.~\ref{subsec:analytic-facet}. We use NV upper bounds on the quantum value to obtain lower bounds on the minimum detection efficiency for the stated measurement class. Conversely, feasible see-saw values give upper bounds on that minimum. When the white-noise contribution is fixed, the same ordering applies to critical visibility.

An open problem raised in Ref.~\cite{divianszky2023certification} is whether retaining the no-click event as a genuine third outcome, and testing the full data against $\mathcal C_2(X,Y,3)$ under the same uniform-loss assumption, can further reduce the detection threshold.

\subsection{Optimal no-click assignments and comparison with white-noise visibility}
\label{subsec:detection_visibility}

Here we connect the two robustness measures introduced above and simplify the calculation of detection thresholds. For a fixed witness and quantum realization, under the uniform-loss assumption, we first determine the optimal output assigned to a no-click event for each measurement input. These choices can be made independently, directly from the witness coefficients, without enumerating all $B^Y$ possible assignments.

We then show that the resulting critical detection efficiency cannot exceed the critical visibility against white noise on the preparations, and identify when the two thresholds coincide. The argument applies to arbitrary POVMs and any number of outcomes. Finally, we discuss optimization over quantum realizations and how the thresholds depend on quantum dimension when the classical message dimension is held fixed.

Consider the functional
\begin{equation}
    I_{\boldsymbol c}[p]=\sum_{b,x,y} c_{bxy}\,p(b|x,y),
\end{equation}
with classical bound \(C\), and let \(Q>C\) denote the quantum value attained by a given quantum strategy. Define
\begin{equation}
    \Gamma_{by}:=\sum_x c_{bxy}.
\end{equation}

Suppose that Bob's detector clicks with probability \(\eta\), independently of the inputs, and that a no-click event for input \(y\) is deterministically assigned to an outcome \(b_y\). The resulting value of the witness is
\begin{equation}
    I_\eta
    =
    \eta Q+(1-\eta)Z_{\mathbf b},
    \qquad
    Z_{\mathbf b}:=\sum_y\Gamma_{b_y y}.
\end{equation}
The corresponding critical efficiency is therefore
\begin{equation}
    \eta_{\mathrm{crit}}(\mathbf b)
    =
    \frac{C-Z_{\mathbf b}}{Q-Z_{\mathbf b}}.
    \label{eq:eta_crit_assignment}
\end{equation}
Since the function
\begin{equation}
    f(z):=\frac{C-z}{Q-z}
\end{equation}
satisfies
\begin{equation}
    f'(z)=\frac{C-Q}{(Q-z)^2}<0,
\end{equation}
minimizing the critical efficiency is equivalent to maximizing
\(Z_{\mathbf b}\). Moreover, the optimization factorizes over Bob's
inputs, yielding
\begin{equation}
    b_y^\star\in\arg\max_b\Gamma_{by},
    \qquad
    Z^\star=\sum_y\max_b\Gamma_{by}.
    \label{eq:optimal_noclick_score}
\end{equation}
Hence, the optimal deterministic assignment can be found directly,
without enumerating all \(B^Y\) possible assignments. The optimum $Z^\star$ is also the one-message classical bound $C_1(\boldsymbol c)$.

We now compare this quantity with the critical visibility against state white noise. Maximally mixed preparations produce
\begin{equation}
 p_{\mathrm{wn}}(b|x,y)=\frac{\Tr M_{b|y}}d=:w_{b|y},
\end{equation}
where $w_{b|y}\geq0$ and $\sum_bw_{b|y}=1$. The corresponding witness value is
\begin{equation}
 W=\sum_y\sum_bw_{b|y}\Gamma_{by}=I_{\mathrm{wn}}.
\end{equation}

The critical visibility \(V_{\mathrm{crit}}\), defined by
\begin{equation}
    V_{\mathrm{crit}}Q+(1-V_{\mathrm{crit}})W=C,
\end{equation}
is
\begin{equation}
    V_{\mathrm{crit}}
    =
    \frac{C-W}{Q-W}.
    \label{eq:white_noise_visibility}
\end{equation}
For every input $y$, the maximum of the numbers $\{\Gamma_{by}\}_b$ is at least their weighted average. Therefore,
\begin{equation}
 Z^\star=\sum_y\max_b\Gamma_{by}
 \geq\sum_y\sum_bw_{b|y}\Gamma_{by}=W.
\end{equation}

Both \(Z^\star\) and \(W\) are values attained by classical behaviors
independent of Alice's input, and hence satisfy \(Z^\star,W\leq C\).
Using the monotonicity of \(f\), we finally obtain
\begin{equation}
    \eta_{\mathrm{crit}}
    =
    \frac{C-Z^\star}{Q-Z^\star}
    \leq
    \frac{C-W}{Q-W}
    =
    V_{\mathrm{crit}}
    .
    \label{eq:eta_leq_visibility}
\end{equation}
Equality holds if and only if, for every $y$, each outcome with $\Tr M_{b|y}>0$ maximizes $\Gamma_{by}$. If every effect has positive trace, this reduces to
\begin{equation}
 \Gamma_{1y}=\Gamma_{2y}=\cdots=\Gamma_{By}
\end{equation}
for every $y$. Randomized no-click assignment cannot improve the result, since it only forms convex combinations of deterministic-assignment scores. The comparison applies to any fixed quantum realization with $Q>C$, including arbitrary POVMs. Optimizing both thresholds over a common realization class preserves the inequality between their infima.

For a fixed witness and classical message dimension $M$, increasing the quantum dimension $d$ cannot increase either optimized threshold when arbitrary measurements are allowed. For detection efficiency this follows by embedding the ideal realization into the larger space. For state white noise, the measurement can be extended on the unused subspace using the optimal deterministic output; the proof is given in Appendix~\ref{app:dimension-thresholds}. This monotonicity applies when the critical visibility is minimized over all allowed preparations and measurements. It need not hold for the visibility of the particular states and measurements chosen to maximize the violation without noise. It also does not compare scenarios in which the classical message dimension increases together with the quantum dimension.

\section{Facets and properties}

Using the combination of enumeration techniques explained in Section~\ref{subsec:enumeration}, we have derived complete and partial lists of facets for a large number of nontrivial scenarios. A classical bounded-message polytope coincides with the full conditional-probability polytope exactly when $M\geq\min\{X,B^Y\}$: Bob has $B^Y$ deterministic response strings, and a deterministic table requires one message per distinct row. Thus the nontriviality condition is $X>M$ and $B^Y>M$. Geometric nontriviality alone does not imply an equal-dimension quantum advantage. As supplementary material~\cite{data_availability}, we will provide files with one representative per facet class for each characterized scenario, with complete, conjecturally complete, or partial status indicated.

When presenting the facet catalogues, we fix the message cardinality $M$ and the number of outcomes $B$. We then abbreviate each scenario by $(X,Y)$, where $X$ is the number of preparations and $Y$ the number of measurement settings.

\subsection{Exact saturation of measurement inputs}
\label{subsec:input-saturation}

For fixed $X$, $M$, and $B$, let $Y_{\mathrm{sat}}(X,M,B)$ be the smallest integer such that every facet of $\mathcal C_M(X,Y,B)$, for arbitrary $Y$, is an input lifting of a facet involving at most $Y_{\mathrm{sat}}$ Bob inputs. An input is essential when its coefficient block cannot be removed using normalization. Shared randomness is allowed, and positivity facets are included in this definition.

\begin{theorem}[Exact input saturation]
\label{thm:ysat}
For $2\le M<X$ and $B\ge2$,
\begin{equation}
Y_{\mathrm{sat}}(X,M,B)=
\begin{cases}
2^{X-1}-1, & M=2,\\[4pt]
\displaystyle\sum_{k=2}^{X-M+1}\binom{X}{k}, & 3\le M<X.
\end{cases}
\label{eq:ysat-main}
\end{equation}
\end{theorem}

Whenever a behavior cannot be simulated with an $M$-valued classical message, some restriction to at most $Y_{\mathrm{sat}}$ of Bob's inputs already witnesses this failure. Conversely, there are facets for which all $Y_{\mathrm{sat}}$ inputs are essential, so no smaller cutoff works uniformly. The cutoff is independent of $B$, although the facet classes themselves generally depend on $B$. For each fixed $X,M,B$, a complete classification through this cutoff therefore suffices for arbitrary $Y$.

\emph{Proof outline.}
For the upper bound, group preparations that send the same deterministic message into message classes. All of Bob's measurements must be compatible with a common probability distribution of Alice's encodings. For $M\ge3$, compatibility can be expressed through the probabilities that specified subsets occur as entire message classes. A class contains at most $X-M+1$ preparations. The singleton probabilities are determined by the others, which obey one further affine constraint, leaving at most $R-1$ free parameters, where $R=\sum_{k=2}^{X-M+1}\binom{X}{k}$. Helly's theorem then implies that any failure of compatibility is already visible in at most $R$ inputs. For $M=2$, the same argument uses the distribution over the $2^{X-1}-1$ nonconstant bipartitions.

To attain the bound, we construct binary tests that distinguish one preparation from a specified set of others. Assigning the distinguished preparation the same message as any preparation in that set reduces Bob's optimized score. Positive weights are chosen so that every deterministic encoding using exactly $M$ nonempty message classes attains the same total optimum. Refining an encoding with fewer classes cannot decrease its optimized score. Ties between optimal responses determine each coefficient column up to scale. The uniqueness of the relative weights then establishes facetness. Separate constructions cover a bit and three or more messages. Finally, splitting a binary outcome preserves facetness for these constructions and gives the same number of essential inputs for every $B\ge2$. The full proof, including the explicit weights, is given in Appendix~\ref{app:ysat}.

\subsection{Number of classes, detection efficiency and resistance to noise}

Table~\ref{tab:catalogues} lists the numbers of nontrivial facet classes, including lifted classes and excluding positivity. The following tables give bounds on the minimum visibility and detection efficiency over each catalogue and the input liftings included in the analysis. The target interval width is $10^{-5}$. For complete catalogues, matching lower and upper bounds determine the scenario minima to the reported precision, subject to the numerical assumptions described in Appendix~\ref{app:numerical-status}. For partial or conjecturally complete catalogues, the conclusions apply to the facets studied. We call an upper endpoint achieved when a corresponding realization is provided. For $\Vmax$, that realization must also attain the exact noiseless maximum. Determining a catalogue minimum does not require determining every individual quantum score.

\subsubsection{Bit communication, binary outcomes $(M=2,B=2)$}

We provide facet catalogues for several $(X,2,Y,2)$ scenarios: (3,2,2,2), (3,2,3,2), (4,2,2,2), (4,2,3,2), (4,2,4,2), (4,2,5,2), (4,2,6,2), (4,2,7,2), (5,2,2,2), (5,2,3,2), (6,2,2,2), (6,2,3,2), (7,2,2,2), and (8,2,2,2), as well as the complete catalogues for (7,2,3,2) and (8,2,3,2). Table~\ref{tab:catalogues} summarizes the class counts and completeness status for selected scenarios. Lists verified by completed exhaustive enumerations are complete. Scenarios marked with an asterisk are conjecturally complete: PANDA has not terminated, and its apparent final stage is computational evidence rather than a completeness certificate. We also provide partial lists for (5,2,4,2), (5,2,5,2), (6,2,4,2), and (9,2,3,2).

Table~\ref{tab:2:2} summarizes detection efficiencies and white-noise visibilities for selected scenarios for qubits, qutrits, and classical trits tested against the classical-bit bound. We distinguish nontrivial binary projective measurements from unrestricted binary measurements, whose linear optimum can also be attained with PVMs, possibly including deterministic settings. Theorem~\ref{thm:ysat} gives $Y_{\mathrm{sat}}(X,2,2)=2^{X-1}-1$ for every $X\ge3$. In particular, $Y_{\mathrm{sat}}=3$ for $X=3$ and $Y_{\mathrm{sat}}=7$ for $X=4$. Thus a complete classification through $(4,2,7,2)$ suffices for arbitrary $Y$; direct enumeration at $Y=8$ is unnecessary. Since some of the catalogues through $Y=7$ are conjecturally complete, the resulting all-$Y$ classification retains that qualification.

Regarding visibilities, we find no analyzed facet with $V_{\mathrm{crit}}^G<1/\sqrt2$. For scenarios with complete facet catalogues, this exclusion is certified when the quantum upper bounds establish $\widetilde q_2\leq\sqrt2\,\widetilde C_2$ for every facet. A known construction beating this threshold lies in $(48,2,24,2)$~\cite{divianszky2023certification}; the stronger visibility near $0.69604$ is associated with the refined $97\times97$ Bell construction~\cite{designolle2026better}, and its $194\times97$ PM realization described above. Our finite searches do not establish the smallest possible scenario. The unrestricted qubit minima in Table~\ref{tab:2:2} agree with the $G$-restricted minima within $10^{-5}$. The dimension-three columns test $C_2$ as well; they must not be read as thresholds against a classical trit.

Regarding detection efficiency, the $2\longrightarrow1$ Random Access Code in $(4,2,2,2)$ gives $1/\sqrt2$, with its well-known CHSH connection~\cite{galvao2002foundations}, while $I_4$ gives $2/3$ in $(4,2,3,2)$. The same threshold follows from the $(3,2,3,2)$ facet listed as class~3 in Table~I of Ref.~\cite{Poderini2020}, whose bit bound, qubit value and optimal no-click score are $5$, $6$ and $3$. The smaller-facet families in Appendix~\ref{app:smaller-facets} achieve $\eta_{\mathrm{crit}}=\eta_{\mathrm{crit}}^G\simeq0.64997835$, $0.64798731$ and $0.64172234$ in $(X,Y)=(5,4),(6,3),(6,4)$, respectively, all below $2/3$.

Guided by their preparation and measurement structures, we construct a facet in $(10,2,3,2)$ with exact qubit value $4\sqrt3+6\sqrt2$ and critical detection efficiency
\begin{equation}
\eta_{\mathrm{crit}}=\eta_{\mathrm{crit}}^G
=\frac{6}{4\sqrt3+6\sqrt2-6}
\simeq0.6373835250.
\end{equation}
The derivation is given in Sec.~\ref{subsec:analytic-facet}. This improves on the achievable detection threshold reported in Ref.~\cite{divianszky2023certification}, while reducing the number of preparations from $70$ to ten and the number of measurements from $70$ to three. It gives a lower bound on $K_D$, rather than on $K_{\mathrm{PM}}=K_G(3)$.

At $(X,Y)=(4,4)$, a qutrit reaches $\eta\simeq0.27792639$ against $C_2$, whereas a classical trit cannot go below $1/3$ under the same loss model. Thus this comparison separates quantum and classical resources of dimension three even though the tested facets bound a bit. The nearby endpoints at larger $Y$ agree within their run uncertainties. These classical bounds are derived in Appendix~\ref{app:classical-controls}.

For three preparations and bit communication, Theorem~\ref{thm:ysat} gives $Y_{\mathrm{sat}}(3,2,2)=3$. Thus the three nontrivial facet classes of $(3,2,3,2)$ generate all nontrivial facet classes for any larger number of Bob inputs by input lifting. In particular, no new classes appear at four or five inputs, although the number of vertices increases from 176 to 736 and 3008, respectively. An analogous saturation holds in bipartite Bell scenarios: every facet of $(X,Y,A,B)=(3,N,2,2)$ with $N>5$ is an input lifting of a facet of $(3,5,2,2)$. This saturation, observed in earlier facet enumerations~\cite{Jesus2022}, is proved in a companion paper~\cite{BellSaturation}.

\begin{table*}[t]
\centering\small
\setlength{\tabcolsep}{5pt}\renewcommand{\arraystretch}{1.1}
\caption{Binary-outcome thresholds against the classical-bit bound $C_2$. Entries are rounded upper endpoints; $\Delta$ is the largest reported interval width in the row. The minima of $\Vopt$ and $\Vmax$ agree within $10^{-5}$ and share a $V$ column. The qubit minima also agree with the $G$ restriction. Each metric may be attained by a different facet. $\dagger$The qubit efficiency bound at $(5,5)$ includes an input lifting from $(5,4)$; no matching lower bound is reported. The remaining thresholds in that row refer to the analyzed sample. $\ddagger$Includes explicitly imported lifts. Catalogue counts and coverage are reported separately in Table~\ref{tab:catalogues}.}
\label{tab:2:2}
\begin{tabular}{@{}c rr rr rr r@{}}
\toprule
&\multicolumn{2}{c}{Qubit}&\multicolumn{2}{c}{Qutrit}&\multicolumn{2}{c}{Classical trit}&\\
\cmidrule(lr){2-3}\cmidrule(lr){4-5}\cmidrule(lr){6-7}
$(X,Y)$&$V$&$\eta$&$V$&$\eta$&$V$&$\eta$&$\Delta$\\
\midrule
$(3,2)$ & 0.78361162 & 0.70710678 & 0.57142857 & 0.50000000 & 0.57142857 & 0.50000000 & $\gap{9.570}{-14}$\\
$(3,3)$ & 0.76980036 & 0.66666667 & 0.50000000 & 0.33333333 & 0.50000000 & 0.33333333 & $\gap{9.473}{-11}$\\
\addlinespace[2pt]
$(4,2)$ & 0.70710678 & 0.70710678 & 0.57142857 & 0.50000000 & 0.57142857 & 0.50000000 & $\gap{3.215}{-8}$\\
$(4,3)$ & 0.70710678 & 0.66666667 & 0.50000000 & 0.33333333 & 0.50000000 & 0.33333333 & $\gap{8.937}{-10}$\\
$(4,4)$ & 0.70710678 & 0.66666667 & 0.50000000 & 0.27792639 & 0.50000000 & 0.33333333 & $\gap{9.463}{-8}$\\
$(4,5)$ & 0.70710678 & 0.66666667 & 0.50000000 & 0.27792663 & 0.50000000 & 0.33333333 & $\gap{3.326}{-7}$\\
$(4,6)$ & 0.70710678 & 0.66666667 & 0.50000000 & 0.27792656 & 0.50000000 & 0.33333333 & $\gap{2.636}{-7}$\\
$(4,7)$ & 0.70710678 & 0.66666667 & 0.50000000 & 0.27792665 & 0.50000000 & 0.33333333 & $\gap{3.517}{-7}$\\
\addlinespace[2pt]
$(5,3)$ & 0.70710678 & 0.65985969 & 0.50000000 & 0.33333333 & 0.50000000 & 0.33333333 & $\gap{8.638}{-10}$\\
$(5,4)$ & 0.70710678 & 0.64997835 & 0.50000000 & 0.27792658 & 0.50000000 & 0.33333333 & $\gap{2.811}{-7}$\\
$(5,5)^\dagger$ & 0.70710678 & $\le0.64997835$ & 0.50000000 & 0.27792630 & 0.50000000 & 0.33333333 & ---\\
\addlinespace[2pt]
$(6,4)^\ddagger$ & 0.70710678 & 0.64172234 & 0.50000000 & 0.27792630 & 0.50000000 & 0.33333333 & $\gap{1.000}{-5}$\\
$(7,3)$ & 0.70710678 & 0.64798731 & 0.50000000 & 0.33333333 & 0.50000000 & 0.33333333 & $\gap{1.302}{-9}$\\
\bottomrule
\end{tabular}
\end{table*}

\subsubsection{Analytical construction of the $63.74\%$ facet}
\label{subsec:analytic-facet}
Write $E_{xy}=p(+|x,y)-p(-|x,y)$, assigning no-click events
to $+$. Preparations and measurements are indexed from zero.
The smaller facets and their qubit realizations are described in
Appendix~\ref{app:smaller-facets}.

Starting from the $(6,2,3,2)$ facet $F$ in Eq.~\eqref{eq:smaller-F}, split its row
$(2,1,1)$ into $(1,1,0)$ and $(1,0,1)$, reverse the row
$(1,0,-1)$, and add
$(1,1,1)$, $(1,-1,-1)$, and $(0,1,-1)$.
After reordering, this gives
\begin{equation}
\label{eq:central-C}
C=
\begin{pmatrix}
1&1&1\\
1&-1&-1\\
-1&1&-1\\
-1&-1&1\\ \hline
1&1&0\\
1&-1&0\\
1&0&1\\
-1&0&1\\
0&1&1\\
0&1&-1
\end{pmatrix},
\qquad
W=\sum_{x,y}C_{xy}E_{xy}.
\end{equation}
The first four rows form a tetrahedral pattern; the remaining
rows give a sum and a difference for each pair of axes.

To prove the classical bound, let
$\mathbf a,\mathbf b\in\{-1,1\}^3$ be the two decoder strings,
and set
$\mathbf s=(\mathbf a+\mathbf b)/2$ and
$\mathbf t=(\mathbf a-\mathbf b)/2$.
Writing $\mathbf c_x$ for row $x$ of $C$, optimization over
the encoder gives
\begin{equation}
W(\mathbf a,\mathbf b)
=\sum_x\max\{\mathbf c_x\cdot\mathbf a,
                 \mathbf c_x\cdot\mathbf b\}
=2\sum_y s_y+\sum_x|\mathbf c_x\cdot\mathbf t|.
\end{equation}
If $k$ is the Hamming distance between the decoder strings,
directly counting the tetrahedral and pair contributions gives
\begin{equation}
\begin{array}{c|rrrr}
k&0&1&2&3\\ \hline
\sum_x|\mathbf c_x\cdot\mathbf t|&0&8&10&12\\
\max_{\mathbf a,\mathbf b:\,d_H=k}W&6&12&12&12
\end{array}
\end{equation}
and hence $W_{\mathrm C}=12$.
Facetness was checked separately by exact arithmetic: among the $28\,624$ distinct deterministic bit behaviors, $92$ saturate the inequality and have affine rank $29$ in the $30$-dimensional correlator space.

For projective qubit measurements with unit Bloch vectors
$\mathbf m_y$, optimizing the preparations gives
\begin{equation}
W=\sum_x\left\|\sum_y C_{xy}\mathbf m_y\right\|.
\end{equation}
For $\mathbf v_x=\sum_yC_{xy}\mathbf m_y$, the tetrahedral rows satisfy $\sum_{x=0}^3\|\mathbf v_x\|^2=12$, so Cauchy--Schwarz bounds their contribution by
$4\sqrt3$. Each pair contributes at most
\begin{equation}
\|\mathbf m_i+\mathbf m_j\|
+\|\mathbf m_i-\mathbf m_j\|\leq2\sqrt2.
\end{equation}
Three orthogonal measurement directions attain both bounds,
with preparations $\mathbf r_x=\mathbf c_x/\|\mathbf c_x\|$.
Therefore
\begin{equation}
W_{\mathrm Q}=4\sqrt3+6\sqrt2.
\end{equation}
The same value is optimal over all binary qubit POVMs.
Their extremal observables are projective or $\pm I$;
with zero, one, or two projective observables, the respective
upper bounds are $6$, $12$, and $6+6\sqrt2$, all smaller. In the two-projector case, the score is bounded by
\begin{equation}
6+3\bigl(\|\mathbf m+\mathbf n\|+\|\mathbf m-\mathbf n\|\bigr)
\leq6+6\sqrt2,
\end{equation}
where $\mathbf m,\mathbf n$ are the two unit Bloch vectors.

Each column of $C$ sums to $2$, so the optimal deterministic
no-click assignment gives $W_{\emptyset}=6$. Retaining all
trials therefore yields
\begin{equation}
\eta_{\mathrm{crit}}=\eta_{\mathrm{crit}}^G
=\frac{12-6}{4\sqrt3+6\sqrt2-6}
\simeq0.6373835250.
\end{equation}
The row reversal is essential to this construction:
omitting it raises the classical bound to $14$.

Individual row reversals are not classical PM relabelings, and facetness is established by the separate rank calculation above.

This facet gives the exact lower bound
\begin{equation}
K_D\geq\frac{W_{\mathrm Q}-6}{12-6}
=\frac{2\sqrt3}{3}+\sqrt2-1
\simeq1.5689141007.
\label{eq:new-KD-bound}
\end{equation}
For the same realization, the white-noise visibility is
$12/(4\sqrt3+6\sqrt2)\simeq0.7785390720$.
Since all entries of $C$ belong to $\{0,\pm1\}$, the same facet can also be found by exhaustive enumeration of this coefficient family.

\subsubsection{Bit communication, ternary outcomes $(M=2,B=3)$}

For three preparations, Theorem~\ref{thm:ysat} gives $Y_{\mathrm{sat}}(3,2,B)=3$ also for ternary outcomes. No new facet classes can occur beyond three Bob inputs, modulo input lifting. Table~\ref{tab:outcome-comparison} compares binary and ternary outcomes at fixed preparation number, measurement number, transmitted resource, and classical-bit bound.

\begin{table*}[t]
\centering\small
\setlength{\tabcolsep}{7pt}\renewcommand{\arraystretch}{1.1}
\caption{Binary-versus-ternary comparison against $C_2$. Each entry is the rounded upper endpoint of a minimum over the corresponding supplied catalogue. For ternary measurements the visibility column reports general-POVM $\Vopt$; PVM minima and $\Vmax$ agree within $10^{-5}$, not necessarily in every printed digit. Classical noise replaces the transmitted letter by a uniformly random letter. The two noise models and loss are optimized separately.}
\label{tab:outcome-comparison}
\begin{tabular}{clrrrr}
\toprule
&&\multicolumn{2}{c}{Visibility}&\multicolumn{2}{c}{Detection efficiency}\\
\cmidrule(lr){3-4}\cmidrule(lr){5-6}
$(X,Y)$&Resource&$B=2$&$B=3$&$B=2$&$B=3$\\
\midrule
$(3,2)$&Qubit&.78361162&.78361162&.70710678&.70710678\\
&Qutrit&.57142857&.50000000&.50000000&.33333333\\
&Classical trit&.57142857&.50000000&.50000000&.33333333\\
\addlinespace
$(3,3)$&Qubit&.76980036&.76980036&.66666667&.66666667\\
&Qutrit&.50000000&.50000000&.33333333&.33333333\\
&Classical trit&.50000000&.50000000&.33333333&.33333333\\
\addlinespace
$(4,2)$&Qubit&.70710678&.70710678&.70710678&.70710678\\
&Qutrit&.57142857&.49148189&.50000000&.29289322\\
&Classical trit&.57142857&.50000000&.50000000&.33333333\\
\bottomrule
\end{tabular}
\end{table*}

With three preparations, a classical trit can transmit $x$ itself and reproduce every noiseless quantum behavior. The exact loss threshold against a bit is $1/3$ for qutrits and classical trits with two ternary measurements, as proved in Appendix~\ref{app:classical-controls}. The corresponding qutrit visibility minimum is numerically consistent with $1/2$; the letter-noise simulation alone does not prove this for arbitrary depolarized quantum preparations.

With four preparations and two measurements, a third outcome lowers the qutrit thresholds against a bit from $(V,\eta)\simeq(0.57142857,0.50000000)$ to approximately $(0.49148189,0.29289322)$. The visibility and efficiency witnesses are different, with local IDs 240 and 86 in the $(4,2,2,3)$ catalogue. Their achieved noiseless scores are approximately $5.414213562$ and $6.414213562$, exceeding the respective trit bounds 5 and 6. Their thresholds in Table~\ref{tab:outcome-comparison}, however, test the bit bounds 3 and 4. A classical trit cannot improve on $V=1/2$ or $\eta=1/3$ under the respective noise models. Thus the new thresholds also distinguish qutrits from classical trits of the same transmitted dimension.

The qubit minima show no resolved improvement from a third outcome in these three catalogues. This does not imply that POVMs are irrelevant to every facet: for facet 2 of $(3,2,2,3)$, a feasible POVM score $3.25$ exceeds the retained PVM upper bound $3.236067979$. The final minimum intervals for $(4,2,2,3)$ are given in Table~\ref{tab:ternary-bit-intervals}. In particular, its qutrit-POVM $\Vmax$ upper endpoint is $0.4914819062$, whereas the $\Vopt$ endpoint is $0.4914818926$.

\subsubsection{Trit communication, binary outcomes $(M=3,B=2)$}

For four preparations, the exact input-saturation cutoff is $Y_{\mathrm{sat}}(4,3,2)=\binom42=6$. Completed exhaustive enumerations for $(X,Y)=(4,2),(4,3),(4,4),(4,5)$ give respectively 1, 6, 30, and 87 facet classes, including lifted classes and excluding positivity. These counts are listed alongside the numerical bounds in Table~\ref{tab:trit-binary}. The available catalogue status is stated separately in Table~\ref{tab:catalogues}. The qutrit minimum at six settings is not determined here.

\begin{table}[t]
\centering\footnotesize
\setlength{\tabcolsep}{3pt}
\caption{Four preparations and binary outcomes, tested against $C_3$. Class counts include lifted facets and exclude positivity. Rounded upper endpoints are shown. The $V$ column represents agreement of $\Vopt$ and $\Vmax$ within $10^{-5}$; $\Delta$ is the largest interval width retained in the numerical run. $\dagger$The exact facet construction in Appendix~\ref{app:tetrahedral-qutrit} improves the feasible efficiency endpoint to $3/4$; the original numerical endpoints were $0.75000023$ and $0.75000015$. This does not make the catalogue-wide minimum exact. The classical-quart efficiency values also apply to ququarts: on four preparations both resources reproduce every ideal behavior, so their score bounds and detection thresholds coincide. No quantum optimization is required for this column.}
\label{tab:trit-binary}
\begin{tabular}{crrrrr}
\toprule
&&\multicolumn{2}{c}{Qutrit}&Classical quart&\\
\cmidrule(lr){3-4}
$Y$&Classes&$V$&$\eta$&$\eta$&$\Delta$\\
\midrule
2&1&.92705098&.92705098&.75000000&$\gap{2.535}{-9}$\\
3&6&.83377889&.76980036&.66666667&$\gap{6.144}{-8}$\\
4&30&.83377889&$\tfrac34{}^{\dagger}$&.50000000&$\gap{2.336}{-7}$\\
5&87&.83377889&$\tfrac34{}^{\dagger}$&.50000000&$\gap{1.485}{-7}$\\
\bottomrule
\end{tabular}
\end{table}

Table~\ref{tab:trit-binary} gives the qutrit visibility $0.83377889$ at three binary settings and the efficiency $0.76980036$. A fourth setting lowers the efficiency to $3/4$ within the reported numerical uncertainty; an explicit tetrahedral qutrit realization attains $3/4$ exactly for its facet. No fifth-setting improvement is resolved. With four preparations, a classical quart can transmit the label $x$ and reproduce every ideal behavior. Four orthogonal quantum states reproduce the same behaviors, so $C_4=Q_4$ for every witness. Since the optimal no-click score depends only on the witness, their detection thresholds also coincide. The universal loss simulation and matching witnesses in Appendix~\ref{app:classical-controls} give the exact threshold $1/2$ for binary outcomes at every $Y\ge4$. This argument requires no ququart optimization and makes no assumption about quantum white-noise visibility. The larger-$X$ trit rows retained from the earlier study are listed separately in Appendix~\ref{app:numerical-status}.

\subsubsection{Trit communication, ternary outcomes $(M=3,B=3)$}

The $(4,3,3,3)$ refinement uses 1,121 representatives including positivity, hence 1,120 nontrivial classes in the conjecturally complete catalogue analyzed here. Table~\ref{tab:trit-ternary} gives the final qutrit intervals. At fixed $(X,M,Y)=(4,3,3)$, a third outcome lowers the efficiency from $0.76980036$ to approximately $0.76822610$, a resolved decrease of $0.15743$ percentage points. Three-outcome PVMs already attain this gain. The visibility remains near $0.83377889$, with no improvement larger than $10^{-5}$ resolved over the analyzed catalogues.

\begin{table*}[t]
\centering\small
\setlength{\tabcolsep}{8pt}\renewcommand{\arraystretch}{1.1}
\caption{Qutrit minimum intervals for $(4,3,3,3)$ against $C_3$, rounded as retained in the final refinement. Facet IDs are local to this catalogue and identify achieved endpoints, not unique minimizers. The exact strategy of Appendix~\ref{app:trine-qutrit} sharpens all feasible visibility endpoints to $6/(2+3\sqrt3)\simeq0.833778892879991$ without changing the retained numerical lower bounds.}
\label{tab:trit-ternary}
\begin{tabular}{llrrr}
\toprule
Measurements&Metric&Lower&Upper&Facet\\
\midrule
POVM&$\Vmax$&.8337757766&.8337788945&86\\
&$\Vopt$&.8337757766&.8337788929&774\\
&$\eta$&.7682260317&.7682261033&242\\
\addlinespace
PVM&$\Vmax$&.8337787959&.8337788929&86\\
&$\Vopt$&.8337785446&.8337788929&774\\
&$\eta$&.7682260317&.7682261018&242\\
\bottomrule
\end{tabular}
\end{table*}

Two analytical reductions close the remaining competitive efficiency intervals. Classes 139, 166, and 577 obey the all-POVM bound $Q_3\le43/5$, with $C_3=8$ and $C_1=6$, so their efficiency is at least $10/13>0.76822610$. For classes 240, 242, 534, and 535, a projective-sufficiency theorem proves equality of the qutrit POVM and PVM score maxima. The numerical PVM upper bounds therefore also bound the optimal POVM efficiency. The seven facet classes reduce to two forms of score optimization after merging outcomes with identical coefficients. They remain inequivalent under the allowed relabelings. The proofs are in Appendix~\ref{app:efficiency-reductions}.

The new trine witness, facet 86 in this catalogue, has the exact all-POVM qutrit value $2+3\sqrt3$. Every measurement has zero white-noise contribution for this witness, giving
\begin{equation}
 \Vopt=\Vmax=\frac6{2+3\sqrt3}
 \simeq0.833778892879991.
\end{equation}
This is a single-facet equality, not an exact scenario-wide lower bound. Its own efficiency equals the same ratio and does not compete with facet 242.

We find no qubit violation among the facets studied. A classical quart has exact minimum visibility $2/3$ under uniform letter noise and exact minimum detection efficiency $1/2$, as shown in Appendix~\ref{app:classical-controls}. The efficiency threshold also applies to ququarts because $C_4=Q_4$ on four preparations. Thus the efficiency improvement from $2/3$ with three binary settings to $1/2$ with ternary settings is already achievable with a classical quart. Equality of the noiseless score bounds does not by itself establish equality between classical letter-noise visibility and quantum state-white-noise visibility.

\subsection{Facets from the $(7,3,3,3)$ perfect game}

In Ref.~\cite{cruzeiro2026perfect}, we presented a perfect game inequality in scenario $(7,3,3,3)$. We use the winning sets in Eq.~(36) of that reference and its one-based indexing. The perfect-game inequality $I_{\rm pg}[p]\leq20$, with
$Q_3=S=21$, is not facet-defining. A simple conversion is
to remove one of the following three winning terms:
\begin{equation}
 F[p]=I_{\rm pg}[p]-p(b|x,y)\leq19,
\end{equation}
where $(b,x,y)$ is $(3,1,2)$, $(1,2,1)$, or $(2,3,3)$.
Exact enumeration gives the classical trit bound $19$.
The resulting inequalities are equivalent under relabelings.
Each has 472 saturating deterministic behaviors with affine
rank $41=7\cdot3\cdot(3-1)-1$, and therefore defines a facet
of the classical trit polytope. The original qutrit strategy
attains $F=20$, its algebraic maximum. The removed term was the only winning output in its input block, which now has zero coefficients. Thus $F$ scores 20 nonzero question blocks. If all 21 input pairs are retained as a game, declare the zero block an automatic win: this adds a constant one, changes the classical and quantum scores to 20 and 21, and leaves the facet and both thresholds unchanged.

For this realization, the measurements are rank-one qutrit PVMs, so state white noise gives
$p_{\rm wn}(b|x,y)=1/3$ and $F[p_{\rm wn}]=38/3$.
For input-independent detection efficiency, optimally
assigning no-click events to a fixed outcome for each
measurement gives a no-click score of $14$. Consequently,
\begin{align}
 V_{\rm crit}
 &=\frac{19-38/3}{20-38/3}
 =\frac{19}{22}\simeq0.8636,\\
 \eta_{\rm crit}
 &=\frac{19-14}{20-14}
 =\frac56\simeq0.8333.
\end{align}
Thus, this conversion lowers the critical visibility from
$7/8$ to $19/22$, while leaving the optimal detection
threshold unchanged.

\section{Conclusion and future directions}

We have obtained complete and partial facet catalogues for prepare-and-measure scenarios with bit or trit communication and binary or ternary outcomes, and studied their violations by qubits and qutrits. Beyond these finite catalogues, we determine the exact input-saturation number for every $2\le M<X$ and $B\ge2$, and construct facets attaining it. The cutoff is $2^{X-1}-1$ for a bit and $\sum_{k=2}^{X-M+1}\binom{X}{k}$ for three or more messages, independently of the outcome number. For bit communication with binary outcomes and three preparations, the known classification therefore covers arbitrary numbers of measurement inputs. For four preparations in the same setting, a complete classification through seven inputs would suffice; the catalogues marked as conjecturally complete retain that qualification.

For the analyzed facets, we distinguish fully optimized state-white-noise visibility from visibility restricted to exact noiseless maximizers, and study detection efficiency under uniform erasure with deterministic no-click assignment. Small qubit examples already attain detection thresholds below $2/3$. Guided by their preparation and measurement structures, we analytically construct a facet with ten preparations and three binary measurements whose exact threshold is $6/(4\sqrt3+6\sqrt2-6)\simeq0.6373835250$. This improves on the achievable threshold reported for the cited $70\times70$ construction while using substantially fewer preparations and measurements. We also convert a $(7,3,3,3)$ perfect-game inequality into a facet, preserving perfect qutrit performance, lowering the visibility of its stated realization from $7/8$ to $19/22$, and retaining the detection threshold $5/6$.

The outcome comparison depends on the transmitted resource and the tested classical bound. Against a bit, two ternary measurements and four preparations allow qutrit thresholds near $V=0.49148189$ and $\eta=0.29289322$, below the classical-trit values $1/2$ and $1/3$. Against a trit, at four preparations and three measurements the qutrit efficiency improves from $0.76980036$ with binary outcomes to $0.76822610$ with ternary outcomes. Projective measurements attain the latter improvement. No visibility gain larger than $10^{-5}$ is resolved on the analyzed catalogues. These numerical conclusions are supported by exact witness bounds and analytical reductions, but retain the stated catalogue and numerical qualifications.

Open directions include studying the quantum violations, white-noise visibility, and detection efficiency of the facet families attaining the input-saturation cutoff, obtaining complete classifications in the remaining small scenarios, and constructing families with improved detection thresholds. It is also natural to study the trade-off between scenario size and robustness, and to determine whether keeping no-click events as a separate outcome can improve the threshold under the same uniform-loss assumption~\cite{divianszky2023certification}. Increasing the quantum dimension, with the witness and classical message dimension held fixed, cannot worsen the fully optimized efficiency or $\Vopt$ when arbitrary measurements are allowed. This argument does not imply monotonicity of $\Vmax$, whose noiseless maximizing set may change; comparisons in which both dimensions increase also require a separate analysis. Exploiting the work of Pauwels \textit{et al.}~\cite{Pauwels2022,Pauwels2022b}, one could also investigate entanglement-assisted dimension-bounded prepare-and-measure scenarios.

\begin{acknowledgments}
This article is dedicated to the memory of my dear friend, Miguel Enrech Casaleiro Mascarenhas Proença, whose lifelong encouragement is largely responsible for the existence of this article. The author thanks Armin Tavakoli, Jef Pauwels, Nicolas Gisin and Sébastien Designolle for many useful discussions over the years. I also acknowledge funding by FCT/MCTES - Funda\c c\~ao para a Ci\^encia e a Tecnologia (Portugal) - through national funds and when applicable co-funding by EU funds under the project UIDB/50008/2020.

The research phase was conducted by the author over the course of six years and more recently was supported by extended discussions with ChatGPT (5.4, 5.5 and 5.6 Sol). ChatGPT was used to assist with the exploration and refinement of candidate mathematical arguments and manuscript organization and editing. The author verified and revised all the material and takes full responsibility for the content of the manuscript.
\end{acknowledgments}

\bibliography{bibl}

\appendix
\onecolumngrid

\section{Exact saturation of Bob's inputs in classical prepare-and-measure correlations}
\label{app:ysat}

Alice receives a preparation label $x\in[X]$, sends one of at most $M$ classical messages, and Bob chooses an input $y\in[Y]$ and returns $b\in[B]$. They may share randomness. We denote the resulting polytope by $\mathcal C_M(X,Y,B)$, with $[n]=\{1,\ldots,n\}$. A Bob input is \emph{essential} to an inequality if its coefficient block cannot be removed using normalization. For $M\ge2$ the polytope is full-dimensional in normalized coordinates. Indeed, Alice can separate any chosen $x$ from all other preparations with two messages. Bob can then change the output on that message at a single $y$, generating every coordinate direction. Adding an input with zero coefficients is an \emph{input lifting}.

For $M=1$ the behaviors are independent of $x$, and for $M\ge X$ Alice can transmit $x$ itself. In both cases the facets in the relative affine hull come from positivity. For $B=1$ there is a single behavior. We treat the remaining cases below. The corresponding question for bipartite Bell facets is considered in Ref.~\cite{BellSaturation}.

We prove Theorem~\ref{thm:ysat}, with the cutoff defined in Sec.~\ref{subsec:input-saturation}.

The upper bound follows because every Bob input must admit the same distribution of Alice's encodings. To attain the bound we construct a binary-outcome inequality. Each Bob input asks him to distinguish one preparation from a specified set of others. If Alice assigns them the same message, his optimized score decreases. We choose the weights so that every encoding incurs the same total loss, and show that this condition fixes all the weights up to a common factor.

\subsection{Upper bound from a common encoding distribution}
\label{ysat:sec:pm-upper}

Assume $2\le M<X$. A deterministic encoding is specified, up to relabeling of the messages, by a partition $\pi$ of $[X]$ into message classes. Without loss of generality, we restrict to partitions into exactly $M$ nonempty classes, since any encoding using fewer messages can be refined into one using exactly $M$ without changing the resulting correlations.

Let
\begin{equation}
t\coloneqq X-M+1.
\label{ysat:eq:tdef}
\end{equation}
A class can contain at most $t$ preparations, since each of the other $M-1$ classes needs at least one. Let $\mu_\pi$ be the probability of choosing partition $\pi$. For each nonempty $G\subseteq[X]$ with $|G|\le t$, the probability that $G$ occurs as an entire message class is
\begin{equation}
q_G\coloneqq\sum_{\pi:\,G\in\pi}\mu_\pi.
\label{ysat:eq:qGdef}
\end{equation}
The sum runs over the partitions that contain $G$ as one of their classes. Each partition defines an indicator vector $v^\pi$ for its classes. The possible vectors $q$ form the polytope
\begin{equation}
\mathcal P_M\coloneqq\operatorname{conv}\{v^\pi:\pi\text{ has }M\text{ classes}\},
\qquad v_G^\pi\coloneqq\mathbf 1[G\in\pi].
\label{ysat:eq:QM}
\end{equation}
Every $q\in\mathcal P_M$ satisfies
\begin{equation}
\sum_{G:\,x\in G}q_G=1\quad(x\in[X]),
\qquad
\sum_G q_G=M.
\label{ysat:eq:qconstraints}
\end{equation}

For a fixed Bob input $y$, consider nonnegative variables $t^y_{Gb}$ satisfying
\begin{equation}
\sum_b t^y_{Gb}=q_G,
\qquad
p(b|x,y)=\sum_{G:\,x\in G}t^y_{Gb}.
\label{ysat:eq:compat}
\end{equation}
The class probabilities compatible with the observed data at $y$ form the convex set
\begin{equation}
\mathcal P_y(p)\coloneqq\bigl\{q\in\mathcal P_M:\text{there exist }t^y_{Gb}\ge0\text{ satisfying \eqref{ysat:eq:compat}}\bigr\}.
\label{ysat:eq:Qydef}
\end{equation}

\begin{lemma}[Common-encoding criterion]\label{ysat:lem:common-encoding}
A behavior belongs to $\mathcal C_M(X,Y,B)$ if and only if
\begin{equation}
\bigcap_{y=1}^{Y}\mathcal P_y(p)\ne\varnothing.
\label{ysat:eq:common-intersection}
\end{equation}
\end{lemma}
\begin{proof}
Shared randomness selects a deterministic encoding. Averaging over that randomness gives $q_G$, and averaging Bob's response over all occurrences of the same exact class gives $t^y_{Gb}$, proving necessity.

Conversely, choose a common $q$ belonging to all sets $\mathcal P_y(p)$, together with a decomposition $q=\sum_\pi\mu_\pi v^\pi$. For each Bob input $y$, choose nonnegative variables $t^y_{Gb}$ satisfying Eq.~\eqref{ysat:eq:compat}. The shared randomness selects $\pi$. If $x$ belongs to $G\in\pi$, Alice sends its message, and Bob outputs $b$ with probability $t^y_{Gb}/q_G$. Classes with $q_G=0$ occur with probability zero. Equation~\eqref{ysat:eq:compat} then reproduces $p(b|x,y)$.
\end{proof}

For $M\ge3$, define
\begin{equation}
R(X,M)\coloneqq\sum_{k=2}^{t}\binom Xk.
\label{ysat:eq:Rdef}
\end{equation}
There are $X$ variables for classes containing one preparation and $R$ for those containing at least two. The first equations in \eqref{ysat:eq:qconstraints} determine
\begin{equation}
q_{\{x\}}=1-\sum_{\substack{G:\,x\in G\\|G|\ge2}}q_G.
\end{equation}
Summing the first equations in \eqref{ysat:eq:qconstraints} over $x$ counts each class $G$ exactly $|G|$ times. Thus $\sum_G |G|q_G=X$, while the second equation gives $\sum_G q_G=M$. Subtracting and dropping the singleton terms, which have coefficient zero, gives
\begin{equation}
\sum_{2\le|G|\le t}(|G|-1)q_G=X-M.
\label{ysat:eq:qremaining}
\end{equation}
Thus the common class-probability vector has at most $R-1$ free parameters.

Helly's theorem says that in an affine space of dimension $h$, if a finite family of convex sets has empty intersection, then some $h+1$ of them already have empty intersection. Applying it to the sets $\mathcal P_y(p)$, if the full intersection is empty, a subfamily indexed by at most $R$ Bob inputs already has empty intersection. Hence every nonclassical behavior has a nonclassical restriction to at most $R$ Bob inputs. Lifting all facet inequalities of these restricted scenarios gives a complete description of the full polytope, so every facet is an input lifting of a facet using at most $R$ inputs. Therefore
\begin{equation}
Y_{\rm sat}(X,M,B)\le R(X,M),\qquad 3\le M<X.
\label{ysat:eq:upperM}
\end{equation}
Bob's number of outcomes changes the response variables $t^y_{Gb}$ but not the common vector $q$, which is why the cutoff is independent of $B$.

For $M=2$, up to exchanging the two messages there are
\begin{equation}
N\coloneqq2^{X-1}-1
\end{equation}
nonconstant bipartitions. A common mixture of them has at most $N-1$ free probabilities, and the same Helly argument in this probability simplex gives
\begin{equation}
Y_{\rm sat}(X,2,B)\le2^{X-1}-1.
\label{ysat:eq:upperbit}
\end{equation}

\subsection{Tightness for binary outcomes}
\label{ysat:sec:ties}

We first prove the lower bound for $B=2$, using correlators $E_{xy}\coloneqq p(+|x,y)-p(-|x,y)$. These $XY$ coordinates are full-dimensional. In the lower-bound proof, $\pi$ always ranges over partitions into exactly $M$ nonempty classes. The deterministic responses to those encodings generate the entire polytope, by refinement.

For each Bob input $y$, let $c_y=(c_{xy})_{x\in[X]}$ be the coefficient vector and $E_y=(E_{xy})_{x\in[X]}$ the correlator vector. Consider
\begin{equation}
I(E)\coloneqq\sum_y c_y\cdot E_y\le\beta.
\label{ysat:eq:I}
\end{equation}
For a fixed deterministic encoding $\pi$, Bob can optimize his output independently for each measurement setting $y$ and each message class $G\in\pi$. The maximum contribution at setting $y$ is
\begin{equation}
C_y(\pi)\coloneqq\sum_{G\in\pi}\left|\sum_{x\in G}c_{xy}\right|.
\label{ysat:eq:Cy}
\end{equation}
A class $G$ is a \emph{tie} at input $y$ when $\sum_{x\in G}c_{xy}=0$. Either output then gives Bob the same score.

To prove that \eqref{ysat:eq:I} is a facet it is enough to show that the deterministic strategies saturating it satisfy only one independent affine equality, namely $I(E)=\beta$.

\begin{proposition}[Tightness from ties]\label{ysat:prop:ties}
Suppose all $c_y$ are nonzero and the following three conditions are satisfied.
\begin{enumerate}
\item every partition $\pi$ into $M$ classes has the same optimum, $\sum_yC_y(\pi)=\beta$.
\item for each input $y$, if a vector $w=(w_x)_{x\in[X]}\in\mathbb R^X$ satisfies
\begin{equation}
\sum_{x\in G}w_x=0
\end{equation}
for every message class $G$ tied at input $y$, then $w$ is proportional to $c_y$.
\item if $\sum_ya_yC_y(\pi)$ is independent of $\pi$, then all $a_y$ are equal.
\end{enumerate}
Then \eqref{ysat:eq:I} is a facet, and every input $y$ is essential.
\end{proposition}
\begin{proof}
Let
\begin{equation}
H(E)=h_0+\sum_{y=1}^{Y}\sum_{x=1}^{X}w_{xy}E_{xy}=0
\end{equation}
be any affine equality satisfied by all deterministic strategies saturating \eqref{ysat:eq:I}, and write $w_y=(w_{xy})_{x\in[X]}$. At a tied class, flip only Bob's output on that class and keep all other outputs optimal. Both strategies saturate, so subtraction gives $\sum_{x\in G}w_{xy}=0$. Property 2 implies $w_y=a_yc_y$.

Now choose optimal outputs for any encoding $\pi$. Then
\begin{equation}
h_0+\sum_ya_yC_y(\pi)=0
\end{equation}
for every $\pi$. Property 3 gives $a_y=a$ for all $y$, and property 1 gives $h_0=-a\beta$. Hence $H=a(I-\beta)$. The saturating deterministic vertices therefore obey exactly one independent affine equality, so \eqref{ysat:eq:I} is a facet.

Finally choose $x$ with $c_{xy}\ne0$ and an encoding in which $\{x\}$ is a message class. Two deterministic behaviors differing only in Bob's output on that class at setting $y$ have identical data at all other settings but different values of $I$. Thus $y$ is essential.
\end{proof}

We now choose tests for which the tied classes determine the coefficient vector at each setting. This leaves two questions. Can we give every encoding the same optimum? Do the relative weights of the settings follow uniquely from that requirement?

\subsection{Star coefficients and a finite certificate}
\label{ysat:sec:stars}

For $i\in[X]$ and a nonempty subset $L\subseteq[X]\setminus\{i\}$, let $e_j\in\mathbb R^X$ denote the $j$th standard basis vector and define the star vector
\begin{equation}
r^{i,L}\coloneqq-e_i+\sum_{j\in L}e_j.
\label{ysat:eq:star}
\end{equation}

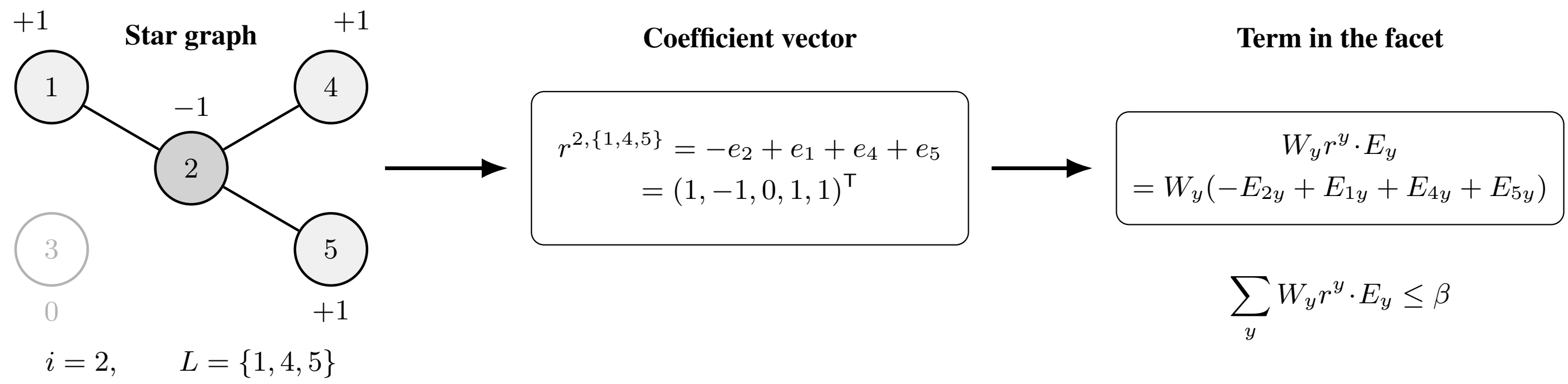
\begin{figure}[H]
\centering
\resizebox{\linewidth}{!}{\begin{tikzpicture}[
    >=Latex,
    every node/.style={font=\small},
    prep/.style={circle,draw=black,thick,minimum size=8mm,inner sep=0pt},
    center/.style={prep,fill=gray!35},
    leaf/.style={prep,fill=gray!12},
    zero/.style={prep,draw=gray!60,text=gray!60},
    eqbox/.style={draw=black,rounded corners,inner sep=5pt,align=center}
]
\node[font=\bfseries] at (0,2.7) {Star graph};
\node[center] (v2) at (0,1.25) {$2$};
\node[leaf]   (v1) at (-1.55,2.15) {$1$};
\node[leaf]   (v4) at (1.55,2.15) {$4$};
\node[leaf]   (v5) at (1.55,0.35) {$5$};
\node[zero]   (v3) at (-1.55,0.35) {$3$};
\draw[thick] (v2)--(v1);
\draw[thick] (v2)--(v4);
\draw[thick] (v2)--(v5);
\node at (0,1.92) {$-1$};
\node at (-1.78,2.88) {$+1$};
\node at (1.78,2.88) {$+1$};
\node at (1.55,-0.34) {$+1$};
\node[text=gray!60] at (-1.55,-0.34) {$0$};
\node at (0,-0.92) {$i=2,\qquad L=\{1,4,5\}$};

\node[font=\bfseries] at (6.2,2.7) {Coefficient vector};
\node[eqbox,minimum width=4.85cm,minimum height=1.7cm] (coefbox) at (6.2,1.25)
{$r^{2,\{1,4,5\}}=-e_2+e_1+e_4+e_5$\\[1mm]
 $=(1,-1,0,1,1)^{\mathsf T}$};

\draw[->,very thick] (2.15,1.25)--([xshift=-2.5mm]coefbox.west);

\node[font=\bfseries] at (12.75,2.7) {Term in the facet};
\node[eqbox,minimum width=4.95cm,minimum height=1.25cm] (facetbox) at (12.75,1.25)
{$W_y r^y\!\cdot\!E_y$\\[1mm]
 $=W_y(-E_{2y}+E_{1y}+E_{4y}+E_{5y})$};

\draw[->,very thick] ([xshift=2.5mm]coefbox.east)--([xshift=-2.5mm]facetbox.west);
\node at (12.75,-0.30) {$\displaystyle \sum_y W_y r^y\!\cdot\!E_y\le\beta$};
\end{tikzpicture}}
\caption{Visualization of a star vector and its contribution to the facet. For $X=5$, $i=2$, and $L=\{1,4,5\}$, the center has coefficient $-1$, the preparations in $L$ have coefficient $+1$, and the remaining preparation has coefficient $0$. Thus $r^{2,\{1,4,5\}}=(1,-1,0,1,1)^{\mathsf T}$. Assigning this star to Bob input $y$ with weight $W_y$ gives the facet term $W_y(-E_{2y}+E_{1y}+E_{4y}+E_{5y})$.}
\label{ysat:fig:star-vector}
\end{figure}
We call $i$ the center of the star and the preparations in $L$ its leaves (the positive set). Associate one star $r^y=r^{i,L}$ with Bob input $y$ and set $c_y=W_yr^y$, where the weights $W_y>0$ will be chosen jointly below. Let $q_y(\pi)\coloneqq1$ when the message class containing $i$ also contains at least one preparation in $L$, and let $q_y(\pi)\coloneqq0$ otherwise. Then
\begin{equation}
C_y(\pi)=W_y\bigl(|L|+1-2q_y(\pi)\bigr).
\label{ysat:eq:star-score}
\end{equation}
Indeed, grouping the coefficient $-1$ at the center with at least one coefficient $+1$ lowers Bob's optimum by exactly two units relative to the sum of absolute coefficients.

The ties are also immediate. Every pair $\{i,j\}$ with $j\in L$ is tied, as is every one-preparation class whose coefficient is zero. Each such class can occur in an encoding with exactly $M$ messages because $M<X$. Therefore a vector $w$ satisfying the tie equations obeys
\begin{equation}
w_k=0\quad(k\notin L\cup\{i\}),
\qquad
w_j=-w_i\quad(j\in L),
\end{equation}
so it is proportional to $r^{i,L}$. Property 2 of Proposition~\ref{ysat:prop:ties} is therefore automatic for every star.

\subsubsection{Exact message classes}
The remaining two conditions of Proposition~\ref{ysat:prop:ties} can be expressed using the same exact message classes already introduced in the upper-bound proof. Let
\begin{equation}
\mathcal G\coloneqq\{G\subseteq[X]:2\le|G|\le t\},
\qquad |\mathcal G|=R(X,M),
\end{equation}
and recall that $v_G^\pi=1$ exactly when $G$ is a message class of the deterministic encoding $\pi$.

For a star $y=(i,L)$ define
\begin{equation}
A_{G,y}\coloneqq\mathbf 1\{i\in G,\ G\cap L\ne\varnothing\},
\qquad G\in\mathcal G.
\label{ysat:eq:Aclass}
\end{equation}
There is exactly one message class containing $i$, so
\begin{equation}
q_y(\pi)=\sum_{G\in\mathcal G}A_{G,y}v_G^\pi.
\label{ysat:eq:qclass}
\end{equation}
Also define $a_G\coloneqq|G|-1$. Equation~\eqref{ysat:eq:qremaining} is precisely
\begin{equation}
\sum_{G\in\mathcal G}a_Gv_G^\pi=X-M
\label{ysat:eq:aclassconstant}
\end{equation}
for every deterministic encoding $\pi$.

We also need the converse. A weight on message classes whose sum is fixed for every encoding must be proportional to $|G|-1$.

\begin{lemma}[Class-weight lemma]\label{ysat:lem:class-weight}
Let $M\ge3$. Suppose numbers $u_G$, indexed by $G\in\mathcal G$, satisfy
\begin{equation}
\sum_{\substack{G\in\pi\\|G|\ge2}}u_G=K
\label{ysat:eq:classweightassumption}
\end{equation}
for every partition $\pi$ of $[X]$ into exactly $M$ message classes. Then there is a scalar $\lambda$ such that
\begin{equation}
u_G=\lambda(|G|-1)
\label{ysat:eq:classweightconclusion}
\end{equation}
for every $G\in\mathcal G$.
\end{lemma}
\begin{proof}
Let $r\coloneqq X-M$, so the allowed nontrivial class sizes are $2,\ldots,r+1$. If $r=1$, every partition has one two-element class and all other classes contain one preparation, so \eqref{ysat:eq:classweightassumption} immediately gives $u_G=K$ for every pair $G$, which is \eqref{ysat:eq:classweightconclusion}.

Assume $r\ge2$. A partition with one class $G$ of size $r+1$ and all remaining classes containing one preparation gives $u_G=K$. Next take two disjoint classes $G,H$ with $|G|=k$ and $|H|=r+2-k$, where $2\le k\le r$. Completing them with one-element classes gives
\begin{equation}
u_G+u_H=K.
\label{ysat:eq:two-classes}
\end{equation}
If two $k$-element subsets $G,G'$ differ in one preparation, then, because $M\ge3$, there is an $(r+2-k)$-element subset $H$ disjoint from both. Applying \eqref{ysat:eq:two-classes} to $(G,H)$ and $(G',H)$ gives $u_G=u_{G'}$. Any two $k$-element subsets can be connected by successive one-element changes, so $u_G$ depends only on $|G|$. Write this value as $u_k$. Thus
\begin{equation}
u_k+u_{r+2-k}=K.
\label{ysat:eq:ukpair}
\end{equation}
Give singletons weight $u_1=0$. For any $2\le k\le r$, choose a partition with one class of size $k+1$ and a singleton. This is possible because the remaining $X-k-2$ preparations can fill the other $M-2$ classes. Move one preparation from the large class to the singleton. Only these two classes change, so their total weight must be the same before and after the move.
\begin{equation}
u_{k+1}+u_1=u_k+u_2.
\end{equation}
Starting from $u_2$, this gives $u_k=(k-1)u_2$ for every $2\le k\le r+1$, as claimed.
\end{proof}

\begin{lemma}[Finite certificate]\label{ysat:lem:certificate}
For $3\le M<X$, choose exactly $R(X,M)$ stars and form the square matrix $A$ of Eq.~\eqref{ysat:eq:Aclass}. Suppose $A$ is nonsingular and there are weights $W_y>0$ and a number $D>0$ such that
\begin{equation}
AW=D a,
\qquad a_G=|G|-1.
\label{ysat:eq:AW}
\end{equation}
Then
\begin{equation}
\sum_yW_yr^y\cdot E_y\le
\beta\coloneqq\sum_yW_y(|L_y|+1)-2(X-M)D
\label{ysat:eq:witness}
\end{equation}
is a facet using all $R(X,M)$ Bob inputs.
\end{lemma}
\begin{proof}
Using \eqref{ysat:eq:star-score} and \eqref{ysat:eq:qclass},
\begin{align}
\sum_y C_y(\pi)
&=\sum_yW_y(|L_y|+1)-2\sum_G(AW)_Gv_G^\pi\\
&=\sum_yW_y(|L_y|+1)-2D\sum_G(|G|-1)v_G^\pi\\
&=\beta,
\end{align}
where the last equality is Eq.~\eqref{ysat:eq:aclassconstant}. Thus every deterministic encoding has the same optimum, which is property 1 of Proposition~\ref{ysat:prop:ties}.

For property 3, suppose $\sum_y\lambda_yC_y(\pi)$ is independent of $\pi$ and set $u_y\coloneqq\lambda_yW_y$. The term $\sum_yu_y(|L_y|+1)$ is already independent of $\pi$, so \eqref{ysat:eq:qclass} implies that
\begin{equation}
\sum_{G\in\mathcal G}(Au)_Gv_G^\pi
\end{equation}
is also independent of $\pi$. The class-weight lemma therefore gives $Au=\gamma a$ for some scalar $\gamma$. Since $AW=Da$ and $A$ is nonsingular,
\begin{equation}
u=\frac{\gamma}{D}W,
\end{equation}
and hence all $\lambda_y$ are equal. Property 2 was established above from the ties. Proposition~\ref{ysat:prop:ties} now proves the claim.
\end{proof}

\subsubsection{A change of basis for the construction}
The matrix $A$ records whether the center shares its message with \emph{at least one} leaf. The construction is easier to verify by asking a second question. Does a star separate its center from \emph{every} other preparation in a specified set $S$? Define
\begin{equation}
T_{S,(i,L)}\coloneqq\mathbf 1\{i\in S,\ S\setminus\{i\}\subseteq L\},
\qquad 2\le|S|\le t,
\label{ysat:eq:T}
\end{equation}
and define the square matrix
\begin{equation}
Z_{G,S}\coloneqq(-1)^{|S|}\mathbf 1\{S\subseteq G\},
\qquad G,S\in\mathcal G.
\label{ysat:eq:Z}
\end{equation}
Then
\begin{equation}
A=ZT,
\qquad
a=Z\mathbf 1.
\label{ysat:eq:AZT}
\end{equation}
Indeed, if $i\notin G$ then $(ZT)_{G,(i,L)}=0$. If $i\in G$, writing $R=S\setminus\{i\}$ gives
\begin{equation}
(ZT)_{G,(i,L)}
=\sum_{\varnothing\ne R\subseteq G\cap L}(-1)^{|R|+1}
=\mathbf 1\{G\cap L\ne\varnothing\},
\end{equation}
which is exactly $A_{G,(i,L)}$. Similarly,
\begin{equation}
(Z\mathbf 1)_G=\sum_{k=2}^{|G|}(-1)^k\binom{|G|}{k}=|G|-1=a_G.
\end{equation}
If subsets are ordered by increasing size, $Z$ is triangular with diagonal entries $(-1)^{|G|}$, and is therefore nonsingular. Consequently, it is enough to construct a square nonsingular $T$ and positive weights satisfying
\begin{equation}
TW=D\mathbf 1.
\label{ysat:eq:TW}
\end{equation}
The left-hand side is the total weight of the tests separating their center from every other preparation in $S$. Thus \eqref{ysat:eq:TW} asks for the same separation score $D$ on every allowed $S$. Equation~\eqref{ysat:eq:AZT} turns this into the required half-loss $D(|G|-1)$ on every class $G$.

\subsection{Positive star weights for three or more messages}
\label{ysat:sec:main}

At $X=M+1$ every relevant class is a pair. Choose the star $(i,\{j\})$ for each $i<j$, with weight one. Each row of $T$ then contains a single one. For larger $X$, a star must also contribute to classes of three or more preparations. We choose the leaves by ordering the preparations. A star at $k$ will contain either all of $1,\ldots,k-1$ as leaves or none of them. This choice lets us compare what happens when we add the earliest preparation to a class.

\begin{lemma}[Positive star weights]\label{ysat:lem:splitting}
For every $3\le M<X$ there is a family of $R(X,M)$ stars for which $T$ is nonsingular and $TW=D\mathbf 1$ with positive integer weights $W_y$ and a positive integer $D$. With the weight of $(1,\{2\})$ set to one, the solution $(W,D)$ is unique.
\end{lemma}
The lemma and the finite certificate give a facet with all $R(X,M)$ inputs. We give the weights here, then prove their balance, positivity, and uniqueness in Sec.~\ref{ysat:sec:weights-proof}.

\subsubsection{How the construction works for three messages}
For $M=3$, start with a table $p_2$ indexed by subsets $U$ of $\{3,\ldots,X\}$. Set $p_2(\varnothing)=-1$ and $p_2(\{3\})=p_2(\{4\})=1$. For every other nonempty $U$ set
\begin{equation}
 p_2(U)=|U\cap\{3,4\}|+3|U\cap\{5,\ldots,X\}|-2.
 \label{ysat:eq:three-start}
\end{equation}
Give $(1,\{2\})$ weight one. For each nonempty $U$ with $p_2(U)>0$, put one star of that weight at each of centers 1 and 2. Its leaves are $U$. Include the other center as an additional leaf precisely when $U$ contains a label at least 5 and $|U|$ is odd.

For $k=3,\ldots,X$, fix the set $K\subseteq\{k+1,\ldots,X\}$ of later leaves. There are two possible tests at $k$. Write $W_k^+(K)$ for the weight of the test with leaves $[k-1]\cup K$, and $W_k^-(K)$ for the weight of the test with leaves $K$. Start from the two table entries with $k$ absent and present. Transfer an amount $\eta_k(K)$ between them, according to
\begin{align}
 W_k^+(K)&=p_{k-1}(K)+\eta_k(K), &
 W_k^-(K)&=p_{k-1}(K\cup\{k\})-\eta_k(K).
 \label{ysat:eq:split-weights}
\end{align}
For three messages, $\eta_3(K)=(-1)^{|K|}$ and $\eta_k(K)=0$ for $k\ge4$. Omit zero weights and the test with no leaves. For the next center, add the two entries,
\begin{equation}
 p_k(K)=p_{k-1}(K)+p_{k-1}(K\cup\{k\}).
 \label{ysat:eq:split-update}
\end{equation}
Thus a table entry represents the combined weight for a given set of later leaves. Only the entry for the empty set may contain an auxiliary signed value. It keeps every table total equal to $D=\sum_U p_2(U)$.

For any $X\ge4$, summing the expression in \eqref{ysat:eq:three-start} over all subsets gives $(3X-14)2^{X-3}$. The corrections at $\varnothing,\{3\},\{4\}$ add $1+2+2$, so $D=5+(3X-14)2^{X-3}$.

For example, take $X=5$. At center 3 the two possible transfers give
\begin{center}
\small
\begin{tabular}{@{}lrrrrr@{}}
\toprule
Later leaves $K$ & $p_2(K)$ & $p_2(K\cup\{3\})$ & $\eta_3(K)$ & $W_3^+(K)$ & $W_3^-(K)$ \\
\midrule
$\{5\}$   & 1 & 2 & $-1$ & 0 & 3 \\
$\{4,5\}$ & 2 & 3 & $+1$ & 3 & 2 \\
\bottomrule
\end{tabular}
\end{center}
For $K=\{5\}$ only the star with leaves $\{5\}$ survives. For $K=\{4,5\}$ the stars with leaves $\{1,2,4,5\}$ and $\{4,5\}$ receive weights 3 and 2. The initial table sums to $D=9$. The family has $\binom52+\binom53=20$ settings. For the class $\{3,5\}$, the tests centered at 3 contribute $3+3+2=8$ to its half-loss, and the test centered at 5 contributes 1. For $\{3,4,5\}$ the three centers contribute $9+8+1=18$. These are the required half-losses $D(|G|-1)$. Every encoding has class sizes $(3,1,1)$ or $(2,2,1)$, so it loses $2\times18=36$ score units in either case. Equation~\eqref{ysat:eq:witness} gives the facet bound $\beta=93$.

\subsubsection{The weights for arbitrary \texorpdfstring{$M$}{M}}
Let $\delta=M-2$ and $m=M+1$. Separate the labels after 2 into the base labels $\mathcal B=\{3,\ldots,m\}$ and the extra labels $\mathcal E=\{m+1,\ldots,X\}$. Write $s=|U|$ and $e=|U\cap\mathcal E|$. In the following formulas $\binom{x}{j}=x(x-1)\cdots(x-j+1)/j!$ for $j\ge1$, $\binom{x}{0}=1$, and $\binom{x}{j}=0$ for $j<0$. This convention also specifies the few entries with $x<0$.

\emph{Centers 1 and 2.}
Give $(1,\{2\})$ weight one and put $p_2(\varnothing)=-\delta$. For nonempty $U\subseteq\mathcal B$ put $p_2(U)=1$ if $|U|=1$, and zero otherwise. For $U$ containing an extra label put $p_2(U)=g(s,e)$, where
\begin{equation}
 g(s,e)=\binom{s-2}{\delta}
       +e\left[\binom{s-2}{\delta-1}+\binom{s-1}{\delta-1}\right].
 \label{ysat:eq:split-start}
\end{equation}
As above, each nonempty $U$ with positive weight gives one star at each of centers 1 and 2. Include the other center as a leaf precisely when $e\ge1$ and $s\equiv \delta\pmod 2$. Set $D=\sum_U p_2(U)$.

\emph{Centers $k=3,\ldots,X$.}
Use Eqs.~\eqref{ysat:eq:split-weights} and \eqref{ysat:eq:split-update}. With $r=M-k$, $s=|K|$ and $e=|K\cap\mathcal E|$, take
\begin{equation}
 \eta_k(K)=
 \begin{cases}
 (-1)^{s+r}\left[\binom{s-2}{r}+e\binom{s-2}{r-1}\right],& k\le M,\\[3pt]
 0,& k>M.
 \end{cases}
 \label{ysat:eq:split-transfer}
\end{equation}
Transfers occur only at centers $3,\ldots,M$. Once $k>M$, the two test weights are read directly from the previous table.

\subsection{Bit communication}
\label{ysat:sec:bit}
For a bit, an encoding is determined by the preparations that send a different message from a fixed reference preparation. Let $n\coloneqq X-1\ge2$ and label preparations $0,1,\ldots,n$. A nonconstant bit encoding is a nonempty subset $U\subseteq[n]$, the preparations sending the message opposite to preparation $0$. There are $N_n\coloneqq2^n-1$ such encodings.

For $i\in[n]$ and $L\subseteq[n]\setminus\{i\}$, including $L=\varnothing$, define
\begin{equation}
g_{i,L}(U)\coloneqq\mathbf 1[i\in U,\ U\cap L=\varnothing],
\qquad d_n(U)\coloneqq\mathbf 1[U=[n]].
\end{equation}
The actual coefficient columns are
\begin{equation}
r_{i,L}\coloneqq-e_i+e_0+\sum_{j\in L}e_j,
\qquad
r_*\coloneqq-e_0+\sum_{j=1}^ne_j.
\end{equation}
The first star attains its full score $|L|+2$ precisely when $i$ is separated from $0$ and every member of $L$. Otherwise it loses two units. Similarly the second attains its full score only when $U=[n]$. Their optimized scores are therefore $|L|+2g_{i,L}(U)$ and $n-1+2d_n(U)$.

We recursively construct two $N_n\times(N_n-1)$ indicator matrices $G_n^\pm$ and positive vectors $w_n^\pm$ such that
\begin{equation}
G_n^\pm w_n^\pm=\mathbf 1\mp\tau_nd_n,
\qquad \tau_n\coloneqq2^{2-n},
\label{ysat:eq:bitbalance}
\end{equation}
and both $[G_n^\pm\mid d_n]$ are nonsingular. At $n=2$, order the encodings as $\{1\},\{2\},\{1,2\}$. Use $g_{1,\{2\}},g_{2,\{1\}}$ for $+$ and $g_{1,\varnothing},g_{2,\varnothing}$ for $-$. This gives
\begin{equation}
G_2^+=\begin{pmatrix}1&0\\0&1\\0&0\end{pmatrix},
\qquad
G_2^-=\begin{pmatrix}1&0\\0&1\\1&1\end{pmatrix},
\qquad
w_2^\pm=\binom11.
\end{equation}
For the step $n\to n+1$, order the rows as old nonempty $U$, then $U\cup\{n+1\}$ with old $U\ne\varnothing$, then $\{n+1\}$, and set
\begin{equation}
G_{n+1}^\pm=
\begin{pmatrix}
G_n^\pm&G_n^\mp&0&0\\
G_n^\pm&0&\mathbf 1&0\\
0&0&1&1
\end{pmatrix},
\qquad
w_{n+1}^\pm=\frac12
\begin{pmatrix}w_n^\pm\\w_n^\mp\\1\\1\end{pmatrix}.
\label{ysat:eq:bitrec}
\end{equation}
These are admissible star columns. The first block retains the old leaves, the second adds $n+1$ to them, and the last two columns are $g_{n+1,\varnothing}$ and $g_{n+1,[n]}$. Direct multiplication gives \eqref{ysat:eq:bitbalance} with $\tau_{n+1}=\tau_n/2$.

For nonsingularity write $\sigma\in\{+1,-1\}$, $G=G_n^\sigma$ and $H=G_n^{-\sigma}$. Assume the old augmented matrices $[G\mid d_n]$ and $[H\mid d_n]$ are nonsingular, as is immediate at $n=2$. A dependence of $[G_{n+1}^{\sigma}\mid d_{n+1}]$ with coefficient blocks $(u,v,a,b,c)$ gives
\begin{equation}
 Gu+Hv=0,\qquad Gu+a\mathbf 1+c\,d_n=0,\qquad a+b=0 .
\end{equation}
Substituting $\mathbf 1=G w_n^{\sigma}+\sigma\tau_n d_n$ in the middle equation yields $u=-a w_n^{\sigma}$ and $c=-a\sigma\tau_n$. The first equation now reads
\begin{equation}
 H(v-a w_n^{-\sigma})=-2a\sigma\tau_n d_n .
\end{equation}
Independence of the columns of $[H\mid d_n]$ and $\tau_n>0$ give $a=0$, then all other coefficients vanish. This completes the induction.

Finally take $K=[G_n^+\mid d_n]$ and the positive vector $\mu=(w_n^+,\tau_n)$, so $K\mu=\mathbf 1$. Up to a setting-dependent constant, the optimized score of each star is twice its column in $K$. Thus the total optimized score with weights $\mu$ is constant over all encodings. Any rescaling $(a_y)$ with the same property satisfies $K(a_y\mu_y)_y=\lambda\mathbf 1$. Since $K$ is nonsingular, $a_y=\lambda$ for every $y$. The star ties give the remaining condition of Proposition~\ref{ysat:prop:ties}. This proves the bit lower bound with all $2^{X-1}-1$ inputs essential.

\subsection{More than two Bob outcomes}
\label{ysat:sec:outcomes}
For $B>2$, split the binary $+$ outcome into $B-1$ outcomes carrying the same coefficients. Merging them maps every classical behavior to a binary classical behavior, and every binary strategy is available using just one copy of $+$. The lifted inequality therefore has the same classical bound and uses the same inputs.

To prove it is a facet, consider any affine equality satisfied by the saturating behaviors. At a fixed input $y$, compare a new copy of $+$ with a reference copy. On every class tied for the binary star, either copy is optimal. Flipping between them gives zero sum on that class for the difference of their coefficient vectors. The star ties force this difference to be proportional to $c_y$. Choose a preparation $x$ with $c_{xy}>0$, which exists for every star used here. The class $\{x\}$ occurs in some encoding, and all copies of $+$ are optimal there. Flipping on this class gives zero difference at $x$, so the proportionality constant is zero. The coefficients of all copies therefore agree at each $y$. The equality descends to the original binary facet, where there is only one independent affine equality apart from normalization. This proves tightness for every $B\ge2$.

\subsection{Proof of the positive-weight lemma}
\label{ysat:sec:weights-proof}
\subsubsection{Why all message classes have the required score}
Consider any weights on stars having all or none of the earlier preparations as leaves. At center $i$, let $p_i(V)$ be the total weight with later leaf set exactly $V$, for $V\ne\varnothing$. Complete the table by setting $p_i(\varnothing)=D-\sum_{V\ne\varnothing}p_i(V)$. At $k\ge2$ use the auxiliary value $W_k^-(\varnothing)=p_k(\varnothing)-W_k^+(\varnothing)$. These definitions apply to any candidate solution, including when we later test uniqueness.

Compare centers $i$ and $k=i+1$ on the same set of labels $\{k,\ldots,X\}$. Make a second table by putting the two kinds of star at $k$ into the entries without and with label $k$, respectively, as follows.
\begin{equation}
 \widetilde p_i(K)=W_k^+(K),\qquad
 \widetilde p_i(K\cup\{k\})=W_k^-(K),\qquad K\subseteq\{k+1,\ldots,X\}.
 \label{ysat:eq:split-compare}
\end{equation}
Then $TW=D\mathbf 1$ is equivalent to
\begin{equation}
 \sum_{V\supseteq U}p_i(V)=\sum_{V\supseteq U}\widetilde p_i(V)
 \qquad\bigl(U\subseteq\{i+1,\ldots,X\},\ |U|\le t-1\bigr)
 \label{ysat:eq:suffix-sums}
\end{equation}
for each $i<X$. Here a sum over supersets counts every test whose later leaves contain the specified labels.

To see the equivalence, take a set $J$ with earliest member $j$, and insert $i<j$. In the separation score $TW$, only a test centered at $i$ can start contributing. Only a test centered at $j$ can stop contributing. Indeed, any contributing center later than $j$ already contains $j$ as a leaf and hence contains all earlier preparations. For $2\le |J|\le t-1$, balance requires the weight at $i$ whose leaves contain $J$ to equal the minus weight at $j$ whose leaves contain $J\setminus\{j\}$. For $J=\{j\}$, the pair score is $D$ when the weight at $i$ whose leaves contain $j$, plus the total plus weight at $j$, equals $D$. In either case the required value is the same for every $i<j$. Equating consecutive values when $i+1<j$, and using the stated value when $i+1=j$, gives \eqref{ysat:eq:suffix-sums}. The empty-set equation follows from the common table total. Conversely, the consecutive comparisons recover the pair scores, then the scores on larger sets by successive insertions.

There is a direct way to check all these sums. For the difference $f_i=p_i-\widetilde p_i$ on the $b=X-i$ later labels, write
\begin{equation}
 f_i(V)=(-1)^{|V|}P_i(V).
 \label{ysat:eq:split-degree}
\end{equation}
Regard $P_i$ as the unique multilinear polynomial in the indicators $\mathbf 1[j\in V]$ giving those values. Then \eqref{ysat:eq:suffix-sums} is equivalent to $\deg P_i\le M-i-1$. A negative degree means $P_i=0$. Indeed, when $P_i$ is expanded in the missing-label indicators $\mathbf 1[j\notin V]$, the coefficient indexed by $A$ is
\begin{equation}
 (-1)^{b+|A|}\sum_{V\supseteq\{i+1,\ldots,X\}\setminus A} f_i(V).
\end{equation}
Thus Eq.~\eqref{ysat:eq:suffix-sums} sets to zero precisely the coefficients of degree at least $b-(t-1)=M-i$. Equivalently, in each required sum a monomial of smaller degree misses a free label, and the two signs for that label cancel.

For the constructed weights at $k=i+1\ge3$, Eq.~\eqref{ysat:eq:split-weights} makes $f_i$ equal to $-\eta_k(K)$ when $k$ is absent and $+\eta_k(K)$ when $k$ is present. Thus $P_i$ is independent of the bit for $k$. For $r\ge0$ it equals $(-1)^{r+1}$ times the bracket in \eqref{ysat:eq:split-transfer}, a polynomial of degree at most $r=M-i-1$. For $r<0$ it vanishes. At the first center the same check gives
\begin{equation}
 P_1(V)=(-1)^{\delta+1}g\bigl(|V\setminus\{2\}|,|V\cap\mathcal E|\bigr).
\end{equation}
For a tail containing an extra label the sign comes from the rule for including the other center. On a nonempty base tail, $g(1,0)=(-1)^\delta$ and $g(s,0)=0$ for $2\le s\le \delta+1$. On the empty tail, $g(0,0)=(-1)^\delta(\delta+1)$, while $p_1(\varnothing)=-(\delta+1)$ and $p_2(\varnothing)=-\delta$. These values verify both choices of the bit for 2. Since $\deg P_1\le \delta=M-2$, we have proved \eqref{ysat:eq:TW} at every center.

\subsubsection{Positive weights and the number of stars}
The stars whose center and leaves all belong to $[m]$ are exactly the pairs $(i,\{j\})$ with $i<j\le m$. Each has weight one. For the general family, the following table specifies which actual weights at $k\ge3$ are positive. The interval $\{k+1,\ldots,m\}$ is empty for $k\ge m$.
\begin{center}
\small
\begin{tabular}{@{}lll@{}}
\toprule
Later leaves $K$ & Leaves $K$ & Leaves $[k-1]\cup K$\\
\midrule
$K=\varnothing$ & no test & $k>m$ \\
$\varnothing\ne K\subseteq\{k+1,\ldots,m\}$ & $|K|=1$ & none \\
$K\cap\mathcal E\ne\varnothing$ & $|K|\ge M-k$ & $|K|\ge M+2-k$ \\
\bottomrule
\end{tabular}
\end{center}

For the starting table, Pascal's identity gives
\begin{equation}
 g(s,e)=\binom{s}{\delta}
 +(e-1)\left[\binom{s-2}{\delta-1}+\binom{s-1}{\delta-1}\right].
\end{equation}
For $e\ge1$ this vanishes for $s<\delta$ and is positive for $s\ge \delta$. At $s=1$ necessarily $e=1$. The $\delta+1$ base singletons and the empty entry already contribute $\delta+1-\delta=1$ to $D$.

For $3\le k\le M$, put $h=k-3$ and $r=M-k$, so $\delta=h+r+1$. Equation~\eqref{ysat:eq:split-update} means that each current entry sums the starting weights over all choices of the $h$ earlier base labels. If $K$ contains base labels only, the absent- and present-bit entries are $(-(r+1),1)$ for $K=\varnothing$ and $(1,0)$ for $|K|=1$. The transfers $r+1$ and $-1$ give the first two rows of the table. For $2\le |K|\le r+1$ both entries and the transfer vanish.

Suppose $K$ contains an extra label and write $H=\binom{s-2}{r}+e\binom{s-2}{r-1}$. For $s<r$ both entries and $H$ vanish. At $s=r$ the entries are $(0,e)$ and $H=0$. At $s=r+1$ the absent-bit entry is $e$ and the transfer is $-e$, so only the minus weight is positive. These threshold values follow by including enough of the $h$ earlier labels to reach $\delta$ in \eqref{ysat:eq:split-start}. At $s=1$, use $e=1$ and Pascal's identity.

For $s\ge r+2$ both entries exceed $H\ge0$. The absent-bit entry includes $g(s+h,e)$, from selecting all $h$ earlier labels, and the present-bit entry includes $g(s+h+1,e)\ge g(s+h,e)$. Counting only the choices that contain all $h$ labels in each binomial of \eqref{ysat:eq:split-start} gives
\begin{equation}
 g(s+h,e)-H\ \ge\ \binom{s-2}{r+1}+(2e-1)\binom{s-2}{r}>0 .
\end{equation}
This proves the last row for $k\le M$. For $k\ge m$ the transfer is zero. Every nonempty later $K$ contains an extra label, and including the earlier base labels gives a positive term in each entry. At $K=\varnothing$, the plus weight is zero for $k=m$, since $\delta$ base singletons cancel $-\delta$. For $k>m$, all $\delta+1$ base singletons contribute, so this weight is positive as well.

To count the stars, start at $X=m$, with one per pair. When we add a largest label $q+1>m$, the new stars correspond one-to-one to the proper subsets of $[q]$ of size at least $\delta$. To see this, write $J=\{3,\ldots,q\}$. A new star at center 1 or 2 corresponds to its center together with the part of its tail in $J$. These are exactly the subsets containing exactly one of $1,2$ and having size at least $\delta$. For $3\le k\le q$, a new star has $q+1\in K$. Put
\begin{equation}
 V=\{3,\ldots,k-1\}\cup(K\setminus\{q+1\})\subsetneq J.
\end{equation}
The first missing label of $J$ recovers $k$. Map the plus star to $V$ and the minus star to $V\cup\{1,2\}$. The two thresholds in the table say in each case that the image has size at least $\delta$. Finally the star at the new center corresponds to $J$ itself. This exhausts the proper subsets, since $J\cup\{1,2\}=[q]$. Hence the increase is
\begin{equation}
 \sum_{a=\delta}^{q-1}\binom qa
 =\sum_{a=1}^{q-\delta}\binom qa
 =R(q+1,M)-R(q,M),
\end{equation}
where the last equality is Pascal's identity. The family has exactly $R(X,M)$ stars.

\subsubsection{Why the weights are unique}
Consider any solution $(W,D)$ on the selected stars. It suffices to prove that fixing the weight $c$ of $(1,\{2\})$ fixes every other weight. We use the same degree bound \eqref{ysat:eq:split-degree}. Two elementary facts will be useful. A multilinear polynomial of degree at most $s$ is determined by its values on sets of size at most $s$. If it vanishes on sets of size less than $s$, only monomials of degree $s$ can remain. Indeed, the coefficient indexed by a set is its value minus the coefficients of its proper subsets.

\emph{First fix the base pairs.}
Write $v_{ij}=W_{i,\{j\}}$ for $i<j\le m$ and $a_i=p_i(\varnothing)$. Set every extra-label indicator to zero. On the remaining $b=m-i$ base labels, $P_i$ has degree at most $b-2$. Its alternating sum therefore vanishes on the full cube and when any one bit is fixed. For $k=i+1\le m$, the only physical stars on this restriction are the base pairs. Summing $f_i$ with $j>k$ present, with $k$ present, and with $k$ absent gives, respectively,
\begin{equation}
 v_{ij}=v_{kj},\qquad
 a_k=v_{ik}-\sum_{j=k+1}^{m}v_{kj},\qquad
 a_i=-\sum_{j=k+1}^{m}v_{ij}.
 \label{ysat:eq:split-base-rank}
\end{equation}
The first identity makes $v_{ij}$ depend only on $j$. Comparing the two expressions for $a_k$ at successive centers shows that these values coincide for successive $j$. Thus $v_{ij}=c$ for every base pair and $a_i=-(m-i-1)c$ for $1\le i\le m$. Subtract $c$ times our constructed solution. Every base-pair weight and every completed entry with no extra label is now zero at the base centers.

\emph{Next fix a set of extra labels.}
Let $E\ne\varnothing$ be an exact set of extra labels, $e=|E|$, and assume that the weights and table differences vanish on every proper subset of $E$ at the base centers. Fix all extra bits outside $E$ to zero. On the slice with $E$ present the degree of $P_i$ in the base bits is at most $M-i-1-e$. Indeed, before those $e$ bits are set to one, each surviving monomial must contain them all.

At centers 1 and 2 write $x_B,y_B$ for the weights with tail $B\cup E$, $B\subseteq\mathcal B$, and put $r=\delta-e$. Both vanish for $|B|<r$ by the starting support. Summing $f_1$ over the two values of bit 2 gives $x_B-y_B$. After the sign $(-1)^{|B|+e}$ is removed this is a first difference of $P_1$ in bit 2, hence has degree at most $r-1$. Its values on $|B|<r$ force it to vanish identically. Thus $x_B=y_B=z_B$. On the slice without bit 2 the rule for including the other center gives $P_1=(-1)^{\delta+1}z_B$. Consequently $z_B$ is zero if $r<0$ and is otherwise homogeneous of degree $r$ in the base membership indicators.

\emph{Remove one base label at a time.}
At center $i\ge2$ suppose the table on this slice is homogeneous of degree $s=M-i-e\ge0$. This is what we just proved at $i=2$. Put $k=i+1$. On sets excluding $k$, the child plus weight vanishes when the remaining base set has size at most $s$, by the support table. The parent also vanishes for sizes below $s$. Thus the restriction of $P_i$, of degree at most $s-1$, vanishes identically. At size $s$ the parent must consequently vanish whenever $k$ is absent. Every surviving monomial of the parent contains $k$. For $s=0$ this forces the constant itself to be zero.

For $s\ge1$, the restriction of $P_i$ with $k$ present has degree at most $s-2$, since its restriction with $k$ absent vanished. Both the parent and the child minus weight vanish on remaining base sets of size below $s-1$. Hence this restriction also vanishes. The child table is the parent's polynomial with the factor for $k$ removed from every monomial. Iterating reaches degree zero at $i=M-e$, where the preceding argument forces the final constant to vanish. It follows back through the stripped factors that the original polynomial $z_B$ was zero. All later discrepancies on the slice have negative degree and are zero. The same conclusion follows immediately when $r<0$. This completes the induction on $E$ at the base centers. The center-2 table is zero on every slice, so its total $D$ is zero. Beyond the base centers Eq.~\eqref{ysat:eq:split-degree} simply copies zeros. Thus the original solution was $c$ times the constructed solution.

In particular, taking $D=0$ gives only zero weights, since our constructed solution has $D>0$. The square matrix $T$ is nonsingular. This proves Lemma~\ref{ysat:lem:splitting} and completes the proof of Theorem~\ref{thm:ysat}.

\emph{Integer normalization.} The same argument also determines the determinant. The uniqueness argument used only additions, subtractions, and signs, so it applies over every prime field. Border $T$ with the column $-\mathbf 1$ and the row $(e_{12}^{\mathsf T},0)$, where $e_{12}$ selects the star $(1,\{2\})$. This integer matrix is nonsingular modulo every prime, hence has determinant $\pm1$. Cramer's rule with last right-hand entry one and all others zero gives $|\det T|=D$, and therefore $|\det A|=D$.

\section{Maximum ideal violation and detection efficiency}
\label{app:ideal-loss}

Consider a linear prepare-and-measure witness
\begin{equation}
I[p]=\sum_{b,x,y}c_{bxy}p(b|x,y)\leq C_M,
\end{equation}
and fix the quantum dimension and the allowed measurement class.
Write $Q=I[p]>C_M$ for the score of a violating realization.
Assume uniform erasure with probability $1-\eta$, followed by a fixed
output $b_y$ for each measurement setting. Every trial is retained.
The observed behavior and score are
\begin{align}
p_\eta(b|x,y)&=\eta p(b|x,y)+(1-\eta)\delta_{b,b_y},\\
I[p_\eta]&=\eta Q+(1-\eta)Z_{\mathbf b},
\qquad Z_{\mathbf b}=\sum_{x,y}c_{b_yxy}.
\end{align}
The no-click score depends only on the witness and the assignment,
and satisfies $Z_{\mathbf b}\leq C_M$. For $Z_{\mathbf b}<C_M$,
the critical efficiency is
\begin{equation}
\eta_{\mathrm{crit}}(p,\mathbf b)
=\frac{C_M-Z_{\mathbf b}}{Q-Z_{\mathbf b}}.
\end{equation}
This expression is strictly decreasing in $Q$. Hence any realization
maximizing the ideal witness value also minimizes the detection
threshold for the same witness and assignment. If
$Z_{\mathbf b}=C_M$, the algebraic threshold is zero for any
violating realization, so the strict-decrease statement is not needed.

The assignment can also be optimized independently of the realization.
For $Q>C_M$, the ratio $(C_M-z)/(Q-z)$ decreases with $z$, so
\begin{equation}
b_y^\star\in\arg\max_b\sum_xc_{bxy},\qquad
Z^\star=\sum_y\max_b\sum_xc_{bxy}=C_1.
\end{equation}
If $Q_{\max}>C_M$ is the maximum ideal score within the chosen
quantum class, the optimum detection threshold is therefore
\begin{equation}
\eta_{\mathrm{crit}}^{\mathrm{opt}}
=\frac{C_M-C_1}{Q_{\max}-C_1}.
\end{equation}
The proof applies to any fixed quantum dimension and any number of
outcomes under this uniform-erasure model. A feasible score gives an
achievable threshold; a matching quantum upper bound establishes its
optimality.

The fixed no-click behavior is essential to the argument. In a
standard Bell experiment with losses at both parties, terms involving
local marginals generally also appear. The lossy Bell value is then
not determined by the ideal witness value alone. Likewise, the present
argument does not identify the optimum state-white-noise visibility
when the white-noise contribution is allowed to vary with the measurements.

\subsection{Dependence on the quantum dimension at fixed classical bound}
\label{app:dimension-thresholds}

Fix a witness and its classical bound $C_M$, and allow arbitrary quantum measurements. Embedding a $d$-dimensional realization into dimension $D\geq d$ preserves its ideal score $Q$. Since the optimal no-click score $C_1$ is unchanged, the minimum detection threshold cannot increase.

For state white noise, choose $b_y^\star$ attaining the optimal no-click score and extend the measurements as
\begin{align}
\rho_x^{(D)}&=\rho_x^{(d)}\oplus0,\\
M_{b|y}^{(D)}&=M_{b|y}^{(d)}\oplus\delta_{b,b_y^\star}\mathds{1}_{D-d}.
\end{align}
Writing $W_d$ for the maximally mixed-state score, this extension gives
\begin{equation}
W_D=\frac dD W_d+\left(1-\frac dD\right)C_1\geq W_d.
\end{equation}
For $Q>C_M$, the ratio $(C_M-W)/(Q-W)$ decreases with $W$. Thus the embedded realization has no greater critical visibility, and minimizing over realizations proves that the optimized visibility is nonincreasing with quantum dimension. This argument requires unrestricted measurements on the added subspace. It does not imply monotonicity of the visibility of states and measurements chosen to maximize the violation without noise, or of equal-dimension comparisons in which $C_M$ also changes.

\clearpage
\section{Qubit realizations of the smaller facets}
\label{app:smaller-facets}

We collect the smaller-facet families that motivate the construction in
Sec.~\ref{subsec:analytic-facet}, together with their qubit realizations,
no-click assignments, and relations between coefficient matrices.

\subsection{Two $(5,4)$ facets with $65.0\%$ detection efficiency}
\label{app:facets-65}
The best detection threshold found in the partial $(5,4)$ catalogue is approximately $0.65$.

The two facets (file rows 25350 and 146888) are
\begin{equation}
 W_K=\sum_{x=0}^{4}\sum_{y=0}^{3}K_{xy}p(2|x,y)\leq6,
 \qquad K\in\{A,B\},
\end{equation}
with zero coefficients for outcome $1$ and
\begin{equation}
\setlength{\arraycolsep}{1.5pt}
A=\left(\begin{array}{rrrr}
1&1&1&1\\
1&1&0&0\\
1&-1&0&-1\\
-1&1&-1&0\\
-1&-1&1&1
\end{array}\right),\quad
B=\left(\begin{array}{rrrr}
1&1&1&1\\
1&1&0&-1\\
1&-1&-1&1\\
-1&0&1&1\\
-1&0&0&-1
\end{array}\right).
\end{equation}
In the following examples, preparations and measurements are indexed from zero. For pure qubit preparations and projective measurements, we use
\begin{equation}
M_{2|y}=(\mathds{1}_2+\mathbf m_y\cdot\boldsymbol\sigma)/2,
\end{equation}
so $p(2|x,y)=(1+\mathbf r_x\cdot\mathbf m_y)/2$.
A realization of $A$ is
\begin{equation}
\begin{array}{ll}
\mathbf m_{0,1}=(a,0,\pm b),&
\mathbf m_{2,3}=(0,d,\pm c),\\[2pt]
\mathbf r_{0,4}=(\pm a,d,0)/R,&
\mathbf r_1=(1,0,0),\\[2pt]
\multicolumn{2}{l}{\mathbf r_{2,3}=(0,-d,\pm(2b+c))/T,}
\end{array}
\end{equation}
where the upper sign refers to the first index,
$R=\sqrt{a^2+d^2}$, $T=\sqrt{d^2+(2b+c)^2}$,
$a^2+b^2=c^2+d^2=1$, and
\begin{equation}
\begin{array}{l}
(a,b)\simeq(0.800358,0.599523),\\
(c,d)\simeq(0.396631,0.917978).
\end{array}
\end{equation}
Facet $B$ reorders the measurements and replaces four preparations
by their orthogonal states:
\begin{equation}
\begin{array}{l}
\mathbf m'=(\mathbf m_0,\mathbf m_2,\mathbf m_3,\mathbf m_1),\\
\mathbf r'=(\mathbf r_0,-\mathbf r_3,-\mathbf r_4,-\mathbf r_2,-\mathbf r_1).
\end{array}
\end{equation}
The four flipped coefficient rows sum to zero in each column,
so this transformation preserves the witness value. Replacing a preparation by its orthogonal state is not an ordinary classical PM input or output relabeling; the two facets remain inequivalent under those relabelings. Both realizations give
$Q=2+2R+a+T\simeq7.07702559$.

Assigning every no-click deterministically to outcome $2$ gives
$W_{\rm nc}=4$. With uniform detection efficiency $\eta$ and all trials
retained, $W_\eta=\eta Q+4(1-\eta)$, so violation requires
\begin{equation}
 \eta>\frac{6-4}{Q-4}\simeq0.64997835.
\end{equation}
Uniformly random no-click assignments instead give $W_{\rm nc}=2$ and
require $\eta>4/(Q-2)\simeq0.78786288$. The deterministic assignment
therefore enables the quoted $65.0\%$ threshold.

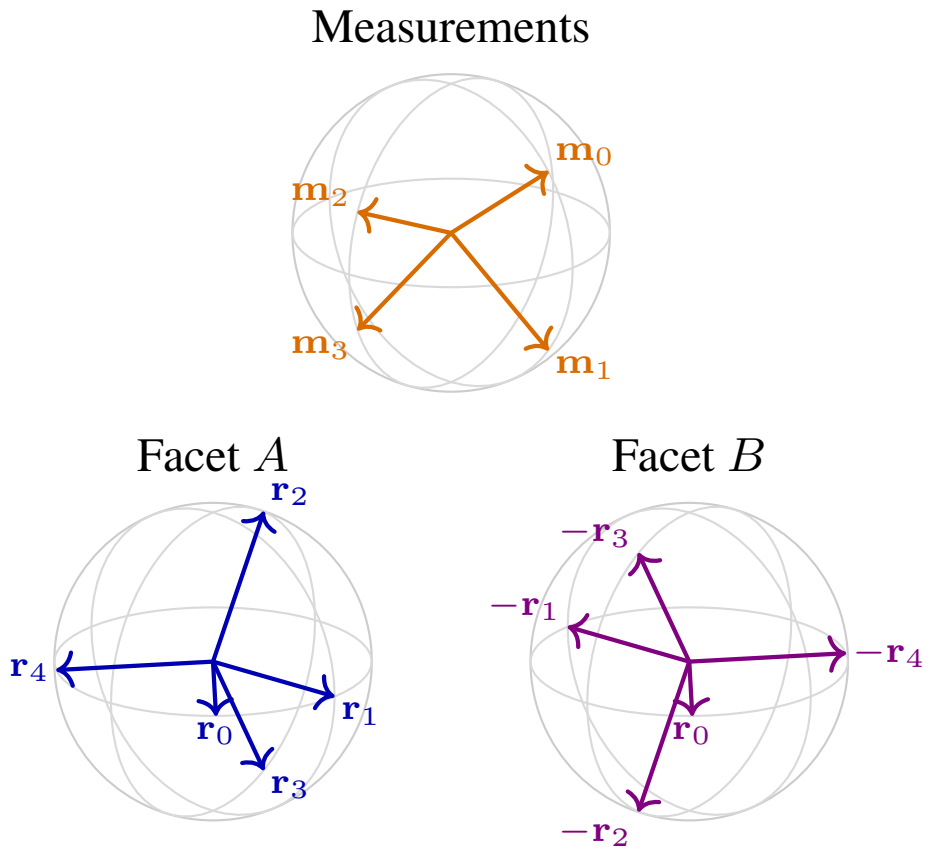
\begin{figure}[H]
\centering
\begin{tikzpicture}[scale=1.05]
\tikzset{pmplot/.style={
 x={(0.766044cm,-0.219846cm)},
 y={(-0.642788cm,-0.262003cm)},
 z={(0cm,0.939693cm)}
}}
\def\pmsphere{\draw[gray!40] (0,0) circle (1);
 \begin{scope}[pmplot]
 \draw[gray!30,domain=0:360,samples=61]
   plot ({cos(\x)},{sin(\x)},0);
 \draw[gray!30,domain=0:360,samples=61]
   plot ({cos(\x)},0,{sin(\x)});
 \draw[gray!30,domain=0:360,samples=61]
   plot (0,{cos(\x)},{sin(\x)});
 \end{scope}
}
\def\pmvector#1#2#3#4#5{\draw[->,thick] (0,0,0)--(#1,#2,#3);
 \node[font=\scriptsize,anchor=#5,inner sep=1pt]
   at (#1,#2,#3) {$#4$};
}
\begin{scope}[shift={(0,2.7)}]
 \node at (0,1.3) {Measurements};
 \pmsphere
 \begin{scope}[pmplot,orange!85!black]
 \pmvector{.800358}{0}{.599523}{\mathbf m_0}{south west}
 \pmvector{.800358}{0}{-.599523}{\mathbf m_1}{north west}
 \pmvector{0}{.917978}{.396631}{\mathbf m_2}{south east}
 \pmvector{0}{.917978}{-.396631}{\mathbf m_3}{north east}
 \end{scope}
\end{scope}
\begin{scope}[shift={(-1.5,0)}]
 \node at (0,1.3) {Facet $A$};
 \pmsphere
 \begin{scope}[pmplot,blue!70!black]
 \pmvector{.657167}{.753745}{0}{\mathbf r_0}{north}
 \pmvector{1}{0}{0}{\mathbf r_1}{north west}
 \pmvector{0}{-.498661}{.866797}{\mathbf r_2}{south west}
 \pmvector{0}{-.498661}{-.866797}{\mathbf r_3}{north west}
 \pmvector{-.657167}{.753745}{0}{\mathbf r_4}{east}
 \end{scope}
\end{scope}
\begin{scope}[shift={(1.5,0)}]
 \node at (0,1.3) {Facet $B$};
 \pmsphere
 \begin{scope}[pmplot,violet]
 \pmvector{.657167}{.753745}{0}{\mathbf r_0}{north}
 \pmvector{0}{.498661}{.866797}{-\mathbf r_3}{south east}
 \pmvector{.657167}{-.753745}{0}{-\mathbf r_4}{west}
 \pmvector{0}{.498661}{-.866797}{-\mathbf r_2}{north east}
 \pmvector{-1}{0}{0}{-\mathbf r_1}{south east}
 \end{scope}
\end{scope}
\end{tikzpicture}
\caption{Shared measurement directions and preparation Bloch vectors.
Four preparations are reversed and reordered; $\mathbf r_0$ is unchanged.}
\end{figure}

\subsection{Five $(6,3)$ facets with $64.8\%$ detection efficiency}
\label{app:facets-648}
The five facet classes $435$, $1219$, $1227$, $2860$, and $3376$ in
the $(6,2,3,2)$ scenario share the threshold
$\eta=\eta^G\simeq64.79873\%$. The reported realizations use rank-one
projective measurements with the same Bloch directions,
\begin{equation}
\begin{aligned}
\mathbf m_0&=(1,0,0),\\
\mathbf m_{1,2}&\simeq(0.13623049,\,0.66069608,\,\pm0.73818828).
\end{aligned}
\end{equation}
These directions are noncoplanar. Let
$(\mathbf r_0,\ldots,\mathbf r_5)$ denote the preparation
vectors in the reported realization of facet $435$. The other strategies follow by reordering
these preparations and replacing selected states by their orthogonal
states:
\begin{equation}
\setlength{\arraycolsep}{3pt}
\begin{array}{c|c|c}
\textnormal{Facet}&\textnormal{Preparations}&
\textnormal{No-click outputs}\\ \hline
435&
(\mathbf r_0,\mathbf r_1,\mathbf r_2,\mathbf r_3,\mathbf r_4,\mathbf r_5)&
(2,2,2)\\
1219&
(\mathbf r_0,-\mathbf r_5,-\mathbf r_4,-\mathbf r_3,-\mathbf r_2,-\mathbf r_1)&
(2,2,2)\\
1227&
(\mathbf r_0,-\mathbf r_5,\mathbf r_2,-\mathbf r_3,\mathbf r_4,-\mathbf r_1)&
(2,2,1)\\
2860&
(\mathbf r_0,\mathbf r_1,\mathbf r_2,-\mathbf r_3,\mathbf r_4,\mathbf r_5)&
(2,1,1)\\
3376&
(\mathbf r_0,-\mathbf r_3,-\mathbf r_2,\mathbf r_4,-\mathbf r_1,\mathbf r_5)&
(1,2,2)
\end{array}
\end{equation}
Here $-\mathbf r_x$ represents the state orthogonal to the pure
state with Bloch vector $\mathbf r_x$; no-click outputs are ordered
by $y=0,1,2$. The five inequalities remain inequivalent under
ordinary input/output relabelings.

For each facet, the classical bound $C$, quantum value $Q$, and
deterministic no-click score $W_{\rm nc}$ satisfy
$Q-C=1.086480218591\ldots$ and $C-W_{\rm nc}=2$.
Retaining all trials therefore gives the common critical efficiency
\begin{equation}
\eta_{\rm crit}
=\frac{C-W_{\rm nc}}{Q-W_{\rm nc}}
=\frac{2}{3.086480218591\ldots}
\simeq0.6479873054.
\end{equation}
Thus the shared threshold follows from the common measurement
geometry, the related preparation ensembles, and the deterministic
no-click assignments listed above.

\subsection{Five $(6,4)$ facets with $64.17\%$ detection efficiency}
\label{app:facets-6417}
The five facet classes $18277$, $146677$, $152485$, $159372$,
and $2072177$ of $(6,2,4,2)$ attain the lowest detection threshold found in the partial catalogue,
$\eta=\eta^G\simeq64.1722\%$.
Their reported qubit realizations share four noncoplanar measurement
bases and six preparation axes, up to relabelings.
Taking class $18277$ as reference, with preparations indexed
$x=0,\ldots,5$, the remaining four ensembles are obtained by
replacing, respectively, the states indexed by
$\{0\}$, $\{5\}$, $\{1,3,5\}$, and $\{3,5\}$ with their
orthogonal partners, $\rho_x\mapsto I-\rho_x$.
Thus the preparation axes coincide, while the oriented ensembles differ.

In correlator notation, all five inequalities have classical
bound $W_{\mathrm C}=8$ and largest qubit value found
$W_{\mathrm Q}\simeq10.233225$, attained with rank-one
projective measurements. We label the outcomes $+$ and $-$ in these representatives. Assigning every no-click event to
$+$ gives $W_{\emptyset}=4$, hence
\begin{equation}
\eta=\eta^G
=\frac{W_{\mathrm C}-W_{\emptyset}}
       {W_{\mathrm Q}-W_{\emptyset}}
\simeq0.641722.
\end{equation}
The deterministic no-click assignment is essential: uniformly
random assignment instead gives a threshold
$8/W_{\mathrm Q}\simeq0.781767$.

\subsection{Relations between the smaller facets}
\label{app:smaller-relations}

For comparison, write $E_{xy}=p(+|x,y)-p(-|x,y)$, assigning no-click events
to $+$. When comparing with the earlier probability witness $A$, we identify $+$ with outcome $2$. Thus its correlator form is $J_A=2W_A-4\leq8$. A representative of the $64.17\%$ family is
\begin{equation}
\label{eq:smaller-L}
L=
\begin{pmatrix}
1&1&1&1\\
0&1&0&-1\\
1&0&-1&0\\
0&0&1&1\\
0&-1&1&-1\\
-1&0&-1&1
\end{pmatrix},
\qquad
\sum_{x,y}L_{xy}E_{xy}\leq8.
\end{equation}
Among the $7\,456$ distinct deterministic bit behaviors, $88$ saturate this inequality and have affine rank $23$ in dimension $24$, verifying facetness.
Indices start at zero. Identifying preparations $1$ and $2$
adds the corresponding rows and recovers the $65.0\%$ facet $A$,
up to relabelings. Identifying measurements $0$ and $1$ instead
adds their columns and gives the $64.8\%$ facet $435$, represented by
\begin{equation}
\label{eq:smaller-F}
F=\begin{pmatrix}
2&1&1\\1&0&-1\\1&-1&0\\0&1&1\\-1&1&-1\\-1&-1&1
\end{pmatrix}.
\end{equation}
In this common correlator normalization, all three representatives have classical bound $8$ and
no-click score $4$. For the displayed representative $F$, the preparations can be reconstructed explicitly as
\begin{equation}
 \mathbf r_x=\frac{\sum_y F_{xy}\mathbf m_y}
 {\|\sum_y F_{xy}\mathbf m_y\|},
\end{equation}
using the measurement directions in Appendix~\ref{app:facets-648}, normalized to unit length after rounding. In the probability normalization of that appendix, $W_F=2+J_F/2$, where $J_F=\sum_{xy}F_{xy}E_{xy}$: the correlator bounds $C_2=8,C_1=4$ become $C_2=6,C_1=4$, and $J_F\simeq10.1729604372$ gives $W_F-C_2\simeq1.0864802186$.

For the displayed representative $L$ in Eq.~\eqref{eq:smaller-L}, a reference measurement tuple is
\begin{equation}
\begin{aligned}
 \mathbf m_0&=(1,0,0),\\
 \mathbf m_1&\simeq(0.89745519,-0.44110564,0),\\
 \mathbf m_2&\simeq(0.21405890,0.66776995,-0.71292502),\\
 \mathbf m_3&\simeq(-0.10244890,-0.69371630,-0.71292491).
\end{aligned}
\end{equation}
Normalize these rounded vectors and take $\mathbf r_x=\sum_yL_{xy}\mathbf m_y/\|\sum_yL_{xy}\mathbf m_y\|$. This physical strategy achieves $J_L\simeq10.2332253840$ and $\eta\simeq0.6417223433$. It specifies a representative of the family, without identifying its coordinate gauge with every stored row. The tuple establishes attainment; the catalogue exclusion bounds are separate.

Thus, allowing the preparations and
measurements to vary independently improves the threshold
by increasing the quantum value.

The $64.8\%$ measurement family
$\mathbf m_0=(1,0,0)$,
$\mathbf m_{1,2}=(u,v,\pm w)$ contains the orthogonal configuration used in Sec.~\ref{subsec:analytic-facet}
at $u=0$ and $v=w=1/\sqrt2$.
Thus the construction combines preparation splitting,
an orthogonal-state substitution, and completion of a
symmetric coefficient pattern within the same measurement
family.

\clearpage
\section{Facet liftings at fixed message dimension}
\label{app:liftings}

We give self-contained proofs of the preparation, measurement-input, and outcome liftings used in Sec.~\ref{subsec:enumeration}. Throughout, $X,Y\ge1$ and the classical message dimension $M\ge2$ is fixed. To describe a split outcome at a single measurement, we allow the number $B_y\ge2$ of outcomes to depend on $y$ and write $P=\mathcal C_M(X,Y,\boldsymbol B)$. The setting of the main text is recovered when $B_y=B$ for every $y$. A deterministic behavior belongs to $P$ exactly when its table of responses has at most $M$ distinct rows.

After eliminating one outcome probability per $(x,y)$, the polytope has dimension $D=X\sum_y(B_y-1)$. Indeed, the constant reference-output table and the tables obtained by changing a single entry are all realizable with two messages and span these coordinates. The same construction, choosing a reference outcome different from the constrained outcome, shows that every positivity inequality $p(b|x,y)\ge0$ defines a facet.

We use the following criterion: a nonempty proper face of a full-dimensional polytope is a facet if every affine equality holding on that face is a multiple of its defining equality. Equalities below are understood within the normalized probability space.

\begin{lemma}
\label{lem:lifting-span}
Let $F$ be a non-positivity facet of $P$. For any fixed input $y$ and outcome $b$, if
\begin{equation}
\sum_{x=1}^X u_x p(b|x,y)=0\qquad\text{for all }p\in F,
\end{equation}
then $u_x=0$ for every $x$.
\end{lemma}
\begin{proof}
The projection of $P$ onto the coordinates $(p(b|x,y))_{x=1}^X$ is the full cube $[0,1]^X$: every binary pattern is realized by sending one of two messages according to whether Bob should return $b$ or a different outcome at $y$. If $u\ne0$, the displayed equality, containing a facet, must define a supporting hyperplane of $P$. Since it passes through the origin of this projected cube, all nonzero $u_x$ must have the same sign. Equality then forces $p(b|x,y)=0$ for every $x$ with $u_x\ne0$. Thus $F$ is contained in a positivity facet and, being itself a facet, equals it. This contradicts the assumption.
\end{proof}

\begin{proposition}[Preparation, input, and outcome lifting]
\label{prop:pm-liftings}
Let $I_{\boldsymbol c}[p]\le C$ define a facet of $P$, with $M\ge2$.
\begin{enumerate}
\item Adding a preparation and assigning zero coefficients to its probabilities preserves facetness.
\item Adding a measurement input and assigning zero coefficients to its probabilities preserves facetness.
\item Splitting an outcome into two outcomes with identical coefficients preserves every non-positivity facet.
\end{enumerate}
The bound $C$ and message dimension remain unchanged. The operations may be repeated, and hence combined.
\end{proposition}

\begin{proof}
Validity follows by deleting the added preparation or input, or by merging the split outcomes: each operation maps a classical behavior in the enlarged scenario to one in the original scenario. Conversely, every old behavior can be extended by duplicating a preparation, adding a constant-output measurement, or assigning zero probability to one of the split outcomes. The lifted inequality therefore has the same tight bound. Write $F$ for the original facet and $F'$ for the lifted equality face. We show that an arbitrary affine equality $H=0$ on $F'$ must be a multiple of the lifted facet equality.

\emph{Preparation lifting.}
Write $H(p,r)=A(p)+U(r)$, where $r$ is the new preparation's row and $U$ is linear in its probabilities. Duplicating any old row $p_i$ gives a behavior in $F'$, so
\begin{equation}
A(p)+U(p_i)=0\qquad(p\in F).
\end{equation}
For $i\ne j$, subtraction gives $U(p_i)-U(p_j)=0$ on $F$. If $U$ is nonconstant on a normalized row, two messages suffice to choose $p_i$ and $p_j$ independently at a maximum and a minimum of $U$. Exchanging the two rows gives both signs of $U(p_i)-U(p_j)$ on $P$. Its zero hyperplane therefore cannot contain the facet $F$. Consequently $U$ is constant on normalized rows and can be absorbed into $A$; the remaining equality is a multiple of $I_{\boldsymbol c}[p]-C=0$. When $X=1$, both the original and the two-preparation polytopes are full products of probability simplices, so the claim follows directly for their positivity facets.

\emph{Measurement-input lifting.}
Positivity facets remain positivity facets. For a non-positivity facet, denote the probabilities at the new input by $r(b|x)$ and choose a reference outcome $b_0$. Using normalization, write
\begin{equation}
H(p,r)=A(p)+\sum_x\sum_{b\ne b_0}h_{bx}\,r(b|x).
\end{equation}
Taking the new measurement always to return $b_0$ gives $A(p)=0$ on $F$. Fix any old input $y$ and outcome $a$. Bob may instead return $b$ at the new input whenever his old response at $y$ would be $a$, and return $b_0$ otherwise. This uses the same encoding and message dimension. Applied to every deterministic strategy saturating the original facet, and then to their convex mixtures, it gives
\begin{equation}
\sum_x h_{bx}p(a|x,y)=0\qquad(p\in F).
\end{equation}
Lemma~\ref{lem:lifting-span} implies $h_{bx}=0$ for every $b,x$. Thus $H=A$ is a multiple of the original facet equality.

\emph{Outcome lifting.}
At a fixed input $y_0$, replace outcome $a$ by $a$ and a new outcome $a'$. If $q$ denotes the enlarged behavior, its merged behavior is
\begin{equation}
p(a|x,y_0)=q(a|x,y_0)+q(a'|x,y_0),
\end{equation}
with all other probabilities unchanged. Setting $q(a'|x,y_0)=0$ embeds $F$ into $F'$. The restriction of $H$ to this embedding is therefore $\lambda(I_{\boldsymbol c}[p]-C)$. For general $q$, we can consequently write
\begin{equation}
H(q)=\lambda(I_{\boldsymbol c}[p]-C)+\sum_x u_x q(a'|x,y_0).
\end{equation}
Relabeling every occurrence of $a$ as $a'$ gives another embedding of $F$ into $F'$, now with $q(a'|x,y_0)=p(a|x,y_0)$. Hence $\sum_xu_xp(a|x,y_0)=0$ on $F$. For a non-positivity facet, Lemma~\ref{lem:lifting-span} gives $u_x=0$, proving the claim.
\end{proof}

The non-positivity qualification is necessary for outcome lifting: splitting the constrained outcome of $p(a|x,y_0)\ge0$ gives $q(a|x,y_0)+q(a'|x,y_0)\ge0$. Its equality face imposes two independent positivity equalities and has codimension two. A positivity facet whose constrained outcome is not split remains a positivity facet.

\clearpage
\section{Catalogue coverage and numerical status}
\label{app:numerical-status}

Table~\ref{tab:catalogues} separates counts from optimization results. Counts include lifted classes and exclude the positivity class. The completeness labels are C for a completed exhaustive enumeration, $*$ for a conjecturally complete catalogue, and Partial for a partial catalogue with a lower-bound count. Numerical optimization does not change any of these labels. Apart from the four-preparation trit/binary files and the positivity convention for the new ternary runs, the counts are retained from the catalogue study and have not been recounted here.

\begin{table}[ht]
\centering\small
\setlength{\tabcolsep}{5pt}\renewcommand{\arraystretch}{1.0}
\caption{Nontrivial facet classes. Scenarios are $(X,M,Y,B)$; each panel fixes $(M,B)$ and lists $(X,Y)$. The trit/binary counts $1,6,30,87$ follow directly from the supplied files with $2,7,31,88$ rows, respectively, after excluding one positivity representative per file. Dashes denote unavailable data, not zero classes.}
\label{tab:catalogues}
\begin{minipage}[t]{.46\textwidth}
\vspace{0pt}\centering
\begin{tabular}{crl}
\toprule
\multicolumn{3}{c}{$(M,B)=(2,2)$}\\
$(X,Y)$&Classes&Status\\
\midrule
$(3,2)$&1&C\\
$(3,3)$&3&C\\
$(4,2)$&2&C\\
$(4,3)$&81&C\\
$(4,4)$&1,389&C\\
$(4,5)$&7,165&C\\
$(4,6)$&14,865&$*$\\
$(4,7)$&17,671&$*$\\
$(5,3)$&3,115&$*$\\
$(5,4)$&$\ge629,867$&Partial\\
$(5,5)$&$\ge36,180$&Partial\\
$(6,3)$&75,814&$*$\\
$(6,4)$&$\ge23\times10^6$&Partial\\
$(7,3)$&952,983&C\\
$(8,3)$&6,226,159&C\\
\midrule
\multicolumn{3}{c}{$(M,B)=(2,3)$}\\
\midrule
$(3,2)$&5&C\\
$(3,3)$&11&C\\
$(4,2)$&584&C\\
$(4,3)$&$\ge9,846$&Partial\\
$(5,2)$&1,395,904&$*$\\
\bottomrule
\end{tabular}
\end{minipage}\hfill
\begin{minipage}[t]{.49\textwidth}
\vspace{0pt}\centering
\begin{tabular}{crl}
\toprule
\multicolumn{3}{c}{$(M,B)=(3,2)$}\\
$(X,Y)$&Classes&Status\\
\midrule
$(4,2)$&1&C\\
$(4,3)$&6&C\\
$(4,4)$&30&C\\
$(4,5)$&87&C\\
$(5,3)$&82&$*$\\
$(6,3)$&4,672&$*$\\
$(7,3)$&$\ge157,951$&Partial\\
\midrule
\multicolumn{3}{c}{$(M,B)=(3,3)$}\\
\midrule
$(4,2)$&20&C\\
$(4,3)$&1,120&$*$\\
$(5,2)$&6,061&$*$\\
$(6,2)$&$\ge4,467,282$&Partial\\
\bottomrule
\end{tabular}
\end{minipage}
\end{table}

The two successive partial counts for $(6,3,2,3)$ were at least 647,130 and 4,467,282; only the latter is shown. These are not to be added. The trit/binary files checked here are \texttt{out\_4322\_matrix}, \texttt{out\_4332\_matrix}, \texttt{out\_4342\_matrix}, and \texttt{out\_4352\_matrix}. Their row lengths are respectively 17, 25, 33, and 41, as required for the $BXY$ probability coefficients and one bound. Each file has one representative per class, so no additional quotient or deduplication is applied.

\subsection{Bounds on minima and their numerical provenance}

The numerical search stops when every unrefined witness is excluded from improving the current minimum by more than $10^{-5}$. For efficiency, an incumbent $\eta_*$ and a checked score bound $Q\le U$ suffice whenever
\begin{equation}
 U\le C_1+\frac{C_M-C_1}{\eta_*-10^{-5}}.
\end{equation}
This criterion permits unresolved individual score intervals. Direct visibility optimization uses noisy-score bounds at fixed visibility. The exact-maximizer quantity additionally requires bounds on the tie set at the noiseless optimum; generic near-maximal strategies are not a substitute for that condition.

Upper score bounds are obtained from the NV relaxations and from the analytical bounds specified in the text, with the intended POVM or PVM measurement model. Imported NV bounds retain their sampled-span provenance and floating-point correction. The resulting intervals are numerical exclusion bounds under those assumptions, rather than exact-arithmetic certificates for every member of a complete catalogue. The interval endpoints in the tables are rounded; a printed decimal equality is not an exact identity.

The source code, exact word sets, solver settings, and per-operation diagnostics will accompany the data release.

For the three bit/ternary catalogues, no class remains capable of improving the reported minima by more than $10^{-5}$ under these numerical assumptions. Seven Bloch--Gram refinements bound the final qubit-POVM loss minimum in $(4,2,2,3)$ with an interval width of $9.529711344\times10^{-6}$.

For $(4,3,3,3)$, the analytical reductions in Appendix~\ref{app:efficiency-reductions}, combined with the numerical bounds, exclude an improvement of the qutrit efficiency minimum larger than $10^{-5}$ over the analyzed catalogue. The transfer from PVM to POVM bounds retains the sampled-span qualifications of the original PVM bounds and applies to score and efficiency, not to visibility.

\begin{table}[ht]
\centering\small
\setlength{\tabcolsep}{7pt}\renewcommand{\arraystretch}{1.0}
\caption{Retained intervals for minima in $(4,2,2,3)$ against $C_2$. These separate the two visibility conventions and both measurement classes. IDs are local to this catalogue. The classical visibility and loss minima $1/2$ and $1/3$ also follow from Appendix~\ref{app:classical-controls}; the rounded quantum intervals do not prove exact algebraic values.}
\label{tab:ternary-bit-intervals}
\begin{tabular}{llrrr}
\toprule
Resource/measurements&Metric&Lower&Upper&Facet\\
\midrule
Qubit POVM&$\Vmax$&.7071057867&.7071067812&102\\
&$\Vopt$&.7071057867&.7071067812&102\\
&$\eta$&.7070972515&.7071067812&102\\
Qubit PVM&$\Vmax$&.7071067595&.7071067812&102\\
&$\Vopt$&.7071067595&.7071067812&102\\
&$\eta$&.7071067401&.7071067812&102\\
Qutrit POVM&$\Vmax$&.4914818854&.4914819062&240\\
&$\Vopt$&.4914818854&.4914818926&240\\
&$\eta$&.2928932141&.2928932189&86\\
Qutrit PVM&$\Vmax$&.4914818668&.4914818942&240\\
&$\Vopt$&.4914818567&.4914818926&240\\
&$\eta$&.2928932026&.2928932188&86\\
Classical trit&$\Vmax$&.5000000000&.5000000000&2\\
&$\Vopt$&.4999999702&.5000000000&2\\
&$\eta$&.3333333333&.3333333333&2\\
\bottomrule
\end{tabular}
\end{table}

\subsection{Additional numerical results}

Table~\ref{tab:legacy-results} gives additional achievable thresholds for which matching lower bounds at $10^{-5}$ are not reported. The visibility is evaluated for the best noiseless realization found numerically; it is not a certified value of $\Vmax$. These entries do not carry the precision claims of Tables~\ref{tab:2:2}--\ref{tab:trit-ternary}.
\begin{table}[ht]
\centering\small
\setlength{\tabcolsep}{7pt}\renewcommand{\arraystretch}{1.0}
\caption{Additional achievable thresholds without matching lower bounds at $10^{-5}$. The resource dimension is $d$, and the tested classical bound is $C_M$. Visibility is evaluated at the best noiseless realization found numerically.}
\label{tab:legacy-results}
\begin{tabular}{cccrr}
\toprule
$(M,B)$&$(X,Y)$&$d$&$V$&$\eta$\\
\midrule
$(2,2)$&$(6,3)$&2&.7071&.6480\\
$(2,2)$&$(8,3)$&2&.7071&.6480\\
$(2,3)$&$(4,3)$&2&.7071&.6667\\
$(3,2)$&$(5,3)$&3&.8065&.7698\\
$(3,2)$&$(6,3)$&3&.8065&.7698\\
$(3,3)$&$(4,2)$&3&.8492&.7836\\
$(3,3)$&$(5,2)$&3&.7925&.6961\\
\bottomrule
\end{tabular}
\end{table}

\clearpage
\section{Classical bounds on visibility and detection efficiency}
\label{app:classical-controls}

\subsection{A noisy classical trit simulated with a bit}

For uniform loss, share a uniformly random trit label $j$. Alice generates the ideal message $m$ and sends the bit $[m=j]$. On a match Bob knows $m$ and samples the ideal response; otherwise he outputs the prescribed no-click value $b_y$. Every message is retained with probability $1/3$. Hence an erased classical trit with $\eta=1/3$ is a classical-bit behavior for every no-click assignment. Mixing with the fallback covers lower efficiencies.

For uniform letter noise, share a uniformly chosen pair of trit labels. Alice sends $m$ if it belongs to this pair and a uniformly chosen member otherwise. The effective message equals $m$ with probability $2/3$ and each other label with probability $1/6$. This is precisely visibility $V=1/2$ plus a uniform trit. Thus no classical trit violates a bit witness at $V\le1/2$ or $\eta\le1/3$ in the respective noise models.

These bounds are attained with three preparations and two ternary measurements. With rows indexed by $x$ and columns by $b$, let
\begin{equation}
 D^{(1)}=\begin{pmatrix}2&0&1\\0&2&1\\1&1&2\end{pmatrix},\qquad
 D^{(2)}=\begin{pmatrix}1&1&0\\1&1&0\\0&0&1\end{pmatrix},
 \qquad S=\sum_{x,b,y}D^{(y)}_{xb}p(b|x,y).
\end{equation}
The two measurements answer the three questions ``is $x=j$?'': the first groups questions 1 and 2 and the second answers question 3. Perfect responses give $C_3=9$, while the maximum column sums give $C_1=6$. With two messages, two preparations must share their response; their target strings differ in two questions, so $C_2=7$. A uniform trit yields $W=5$, giving $V=(7-5)/(9-5)=1/2$ and $\eta=(7-6)/(9-6)=1/3$. The 30 distinct saturating bit behaviors have affine rank 11 in the 12 independent probability coordinates, so this is itself a facet.

For three preparations, every ideal quantum behavior can be reproduced by transmitting $x$ as a trit. The loss simulation therefore also proves a qutrit lower bound $1/3$, attained by three orthogonal states in the displayed witness. This gives the exact qutrit loss minimum for $X=3$, $B=3$, and $Y\ge2$. The classical letter-noise simulation does not establish the same visibility bound for arbitrary depolarized quantum preparations. For binary outputs with $X,Y\ge3$, the correlator matrix $H_{xy}=1-2\delta_{xy}$ on three preparations and settings has $C_1=3$, $C_2=5$, $C_3=9$, and mixed-letter score 1, attaining the same classical thresholds.

\subsection{Four preparation labels simulated with a trit}

Share a uniformly random pair of the four preparation labels. Alice sends one of three messages: first label, second label, or fallback. Bob samples the ideal output when he learns $x$ and uses $b_y$ otherwise. This reproduces every behavior at efficiency $1/2$, for arbitrary settings and outcomes, and mixing covers all lower efficiencies. No resource on four preparations can therefore violate $C_3$ at or below $\eta=1/2$ under this loss model.

Four distinguishable preparations attain the algebraic score 16 of the binary facet in Appendix~\ref{app:tetrahedral-qutrit}, where $C_3=12$ and $C_1=8$. Hence four classical messages or a quantum system of dimension at least four attain $\eta=1/2$ at $Y=4$, and zero-column lifting covers every larger $Y$.

For ternary outputs, group the four questions ``is $x=j$?'' into settings $(1,2)$ and $(3,4)$, each with outputs for the first affirmative answer, the second, or neither. The sum of correct answers over all preparations and questions has $C_1=12$, $C_3=14$, and perfect score 16. Thus $\eta=1/2$ already holds with two ternary settings and persists for $Y\ge2$.

Finally, share a uniformly chosen three-letter subset of four labels, send the ideal label if present, and a uniform member otherwise. Three messages then simulate the four-letter channel with intended-label probability $3/4$ and each other-label probability $1/12$, namely visibility $2/3$. The preceding ternary witness has uniform-letter score 10 and attains $(14-10)/(16-10)=2/3$. This proves the classical-four-letter visibility minimum, without proving the same lower bound for arbitrary four-dimensional depolarized quantum states.

\clearpage
\section{An exact qutrit loss threshold for a binary trit facet}
\label{app:tetrahedral-qutrit}

Let $H_{xy}=1-2\delta_{xy}$ for $x,y\in\{1,2,3,4\}$, and $S=\sum_{xy}H_{xy}E_{xy}$. Each column sums to two, giving $C_1=8$. With three messages two preparations must share a response; their target rows differ at two settings, costing at least four points. Merging exactly one pair attains $C_3=12$, while four messages attain $C_4=16$. The 24 distinct saturating trit behaviors have affine rank 15 in dimension 16, proving facetness.

Take the four unit vectors proportional to $(1,1,1),(1,-1,-1),(-1,1,-1),(-1,-1,1)$, and let $P_x$ be their projectors in $\mathbb C^3$. Preparations $\rho_x=P_x$ and observables $A_y=\I_3-2P_y$ give $E_{xx}=-1$ and $E_{xy}=7/9$ for $x\ne y$. They attain $S=40/3$.

For the upper bound, extremal binary POVMs are projective, including deterministic measurements. Let $k$ be the number of nontrivial settings, $\tau_x=\rho_x-\I_3/3$, and $B_y=A_y-(\Tr A_y)\I_3/3$. Then
\begin{equation}
 S=\frac23\sum_y\Tr A_y+
 \sum_x\Tr\left(\tau_x\sum_y H_{xy}B_y\right).
\end{equation}
A nontrivial qutrit observable has trace $\pm1$ and $\|B_y\|_{\rm HS}^2=8/3$, whereas a deterministic one has trace $\pm3$ and $B_y=0$. Moreover $\|\tau_x\|_{\rm HS}^2\le2/3$ and $H^{\mathsf T}H=4\I_4$. Consequently,
\begin{equation}
 \frac23\sum_y\Tr A_y\le8-\frac{4k}{3},\quad
 \sum_x\|\tau_x\|_{\rm HS}^2\le\frac83,\quad
 \sum_x\left\|\sum_yH_{xy}B_y\right\|_{\rm HS}^2=\frac{32k}{3}.
\end{equation}
Cauchy--Schwarz gives
\begin{equation}
 S\le8-\frac{4k}{3}+\frac{16\sqrt{k}}3
 =\frac{40}{3}-\frac43(2-\sqrt{k})^2\le\frac{40}{3}.
\end{equation}
Thus the all-POVM qutrit optimum is exactly $Q_3=40/3$, and
\begin{equation}
 \eta_{3|3}(H)=\frac{12-8}{40/3-8}=\frac34.
\end{equation}
The mixed-state score of the exhibited strategy is $8/3$, so its visibility is $7/8$, distinct from the lower catalogue visibility near $0.83378$. Input liftings retain its exact loss threshold; this single-facet theorem is not an exact lower bound on the scenario minimum.

\clearpage
\section{An exact trine-facet qutrit bound for ternary measurements}
\label{app:trine-qutrit}
Consider four preparations $\rho_x$ on $\mathbb C^3$, three measurements
$\{E_{b|y}\}_{b=1}^3$, and the score
\begin{equation}S=\sum_{x,y,b} A_{xyb}\Tr(\rho_xE_{b|y}),\end{equation}
with
\begin{equation}
 A_{::1}=0,\qquad
 A_{::2}=\begin{pmatrix}2&2&1\\0&0&-1\\0&-2&1\\-2&0&-1\end{pmatrix},\qquad
 A_{::3}=\begin{pmatrix}2&1&1\\0&1&-1\\0&-1&-1\\-2&-1&1\end{pmatrix}.
\end{equation}
The classical trit bound is $C_3=6$. The coefficient order is $b,x,y$,
with $y$ varying fastest.

\begin{theorem}
For arbitrary qutrit states and arbitrary three-outcome POVMs,
\begin{equation}
 S\le 2+3\sqrt3.
\end{equation}
Rank-one projective measurements attain equality. Consequently,
$Q_3^{\mathrm{POVM}}=Q_3^{\mathrm{PVM}}=2+3\sqrt3$ for this witness.
\end{theorem}

It suffices to prove the bound for pure preparations: with measurements
fixed, each preparation enters the score linearly and can be replaced by
a maximizing pure state. Write $\rho_i=\ket{\psi_i}\bra{\psi_i}$,
$d_{ij}=\sqrt{1-|\langle\psi_i|\psi_j\rangle|^2}$, and let
$D\ge0$ be the determinant of the $3\times3$ Gram matrix of
$\psi_1,\psi_3,\psi_4$. Throughout the proof,
$H_+$ denotes the positive part of a Hermitian matrix and $|H|$ its
operator absolute value.

\begin{lemma}\label{trine:lem:K}
For any three unit vectors $\psi_i,\psi_j,\psi_k$ in $\mathbb C^3$,
with Gram determinant $D$,
\begin{equation}
 K=\rho_k+|\rho_i-\rho_j|\ \succeq\ \frac D2\I.
\end{equation}
\end{lemma}
\begin{proof}
The case $D=0$ follows from positivity. Otherwise, set $d=d_{ij}>0$
and let $P$ project onto $\operatorname{span}\{\psi_i,\psi_j\}$.
The nonzero eigenvalues of $\rho_i-\rho_j$ are $\pm d$, so
$|\rho_i-\rho_j|=dP$. If $t$ is the squared component of $\psi_k$
orthogonal to this plane, then $D=d^2t$ and
$\det(dP+\rho_k)=d^2t=D$.
One eigenvalue of $K$ is $d$. The other two interlace it, and the largest
is at most $1+d\le2$. Thus
\begin{equation}
 \lambda_{\min}(K)
 =\frac{D}{d\lambda_{\max}(K)}\ge \frac D2.
\end{equation}
\end{proof}

We also use the following consequence. For a pure state $\rho$ and
$K\succeq(D/2)\I$, the matrix $\rho-K$ has at most one positive
eigenvalue: on the two-dimensional subspace orthogonal to $\rho$ it is
negative semidefinite. Hence, for every effect $0\preceq B\preceq\I$,
\begin{equation}\label{trine:eq:rankone}
 \Tr[B(\rho-K)]\le\Tr[(\rho-K)_+]\le1-D/2.
\end{equation}
Here $D\le1$ by the Gram determinant bound, so the last upper bound is
nonnegative.

\paragraph{Bounding the three settings.}
For the first setting, the coefficients of outcomes 2 and 3 coincide.
Its contribution satisfies
\begin{equation}
 S_1=2\Tr[(\rho_1-\rho_4)(E_{2|1}+E_{3|1})]\le2d_{14}.
\end{equation}
For the second setting, abbreviate $F_b=E_{b|2}$ and
$R=\rho_1-\rho_3$. Completeness and $\Tr R=0$ give
\begin{equation}
 S_2=\Tr[R(F_2-F_1)]+\Tr[(\rho_2-\rho_4)F_3].
\end{equation}
Since $|R|\pm R\succeq0$ and $F_1,F_2\succeq0$,
\begin{align}
 S_2&\le\Tr[|R|(\I-F_3)]+\Tr[(\rho_2-\rho_4)F_3]\\
 &=2d_{13}+\Tr[F_3(\rho_2-\rho_4-|R|)]
 \le2d_{13}+1-D/2.
\end{align}
The last step uses Lemma~\ref{trine:lem:K} and Eq.~\eqref{trine:eq:rankone}.
For the third setting, write $G_b=E_{b|3}$ and $T=\rho_3-\rho_4$.
Similarly,
\begin{align}
 S_3&=\Tr[T(G_2-G_3)]+\Tr[(\rho_2-\rho_1)G_1]\\
 &\le2d_{34}+\Tr[G_1(\rho_2-\rho_1-|T|)]
 \le2d_{34}+1-D/2.
\end{align}
Therefore every qutrit strategy satisfies
\begin{equation}\label{trine:eq:master}
 S\le2+2(d_{13}+d_{14}+d_{34})-D.
\end{equation}
No projectivity assumption has been made.

\paragraph{Eliminating the overlaps.}
Let $a,b,c$ be the absolute overlaps of the three pairs among
$\psi_1,\psi_3,\psi_4$, and set $s=a^2+b^2+c^2$.
Concavity of the square root and the arithmetic--geometric mean inequality
give
\begin{equation}
 d_{13}+d_{14}+d_{34}\le3\sqrt{1-s/3},\qquad
 D\ge\max\{0,\,1-s-2(s/3)^{3/2}\}.
\end{equation}
The second inequality follows by expanding the Gram determinant and
bounding the real part of the product of its off-diagonal overlaps below
by $-abc$. If $s\ge3/4$, Eq.~\eqref{trine:eq:master} immediately yields
$S\le2+3\sqrt3$. If $0\le s\le3/4$, put $t=\sqrt{s/3}\in[0,1/2]$.
Then
\begin{equation}
 S\le f(t):=1+3t^2+2t^3+6\sqrt{1-t^2}.
\end{equation}
This function is increasing on $[0,1/2]$, since
\begin{equation}
 f'(t)=6t\left(1+t-\frac1{\sqrt{1-t^2}}\right)\ge0,
\end{equation}
as can be checked from
$(1+t)^2(1-t^2)-1=t(2-2t^2-t^3)\ge0$.
Thus $S\le f(1/2)=2+3\sqrt3$, proving the upper bound.

\paragraph{Exact attainment.}
Choose an orthonormal basis $\ket0,\ket1,\ket2$ and the preparations
\begin{equation}
 \ket{\psi_1}=\ket0,\quad \ket{\psi_2}=\ket2,\quad
 \ket{\psi_3}=-\tfrac12\ket0+\tfrac{\sqrt3}{2}\ket1,\quad
 \ket{\psi_4}=-\tfrac12\ket0-\tfrac{\sqrt3}{2}\ket1.
\end{equation}
Let $P=\ket0\bra0+\ket1\bra1$, $F=\ket2\bra2$, and
$H_{ij}=(\rho_i-\rho_j)/(\sqrt3/2)$ for $i,j\in\{1,3,4\}$.
Use the projective measurements, ordered by outcome $b$,
\begin{align}
 (E_{1|1},E_{2|1},E_{3|1})&=((P-H_{14})/2,\ F,\ (P+H_{14})/2),\\
 (E_{1|2},E_{2|2},E_{3|2})&=((P-H_{13})/2,\ (P+H_{13})/2,\ F),\\
 (E_{1|3},E_{2|3},E_{3|3})&=(F,\ (P+H_{34})/2,\ (P-H_{34})/2).
\end{align}
Each row sums to $\I$ and consists of mutually orthogonal rank-one projections. Their contributions are
$S_1=\sqrt3$ and $S_2=S_3=1+\sqrt3$.

\paragraph{Visibility.}
Every coefficient column sums to zero over preparations. Replacing every
state by $\I/3$ therefore gives score zero for any measurements. For this
facet, the optimized and exact-maximizer visibilities coincide:
\begin{equation}
 \Vopt=\Vmax
 =\frac6{2+3\sqrt3}=\frac{18\sqrt3-12}{23}
 \simeq0.833778892879991.
\end{equation}
This theorem concerns one facet; it does not prove that every other facet
has a larger threshold. The existing catalogue bounds still carry their
stated numerical and sampled-span qualifications.

\clearpage
\section{Analytical reductions for the remaining ternary efficiency classes}
\label{app:efficiency-reductions}

All facet IDs in this appendix refer to $(4,3,3,3)$. They are distinct classes under the usual input/output relabelings. Some have duplicate outcome-coefficient columns; merging such outcomes preserves the score, and adding a zero effect reverses the merge for either POVMs or PVMs. This reduces their score optimizations to two effective witnesses. With rows $x$ and columns $y$, the two representatives have $A_{::1}=0$ and
\begin{equation}
 A^{(577)}_{::2}=\begin{pmatrix}2&1&1\\0&1&-1\\0&-1&1\\-2&1&1\end{pmatrix},\qquad A^{(577)}_{::3}=\begin{pmatrix}1&0&0\\1&0&0\\1&0&0\\-1&0&0\end{pmatrix}.
\end{equation}
\begin{equation}
 A^{(242)}_{::2}=\begin{pmatrix}2&2&1\\2&0&-1\\2&-2&1\\-2&0&0\end{pmatrix},\qquad A^{(242)}_{::3}=\begin{pmatrix}0&1&0\\0&1&0\\0&-1&0\\0&0&0\end{pmatrix}.
\end{equation}

The first has $C_3=8$ and the second $C_3=9$; both have $C_1=6$. The duplicate pairs that give binary settings are
\begin{center}\small
\begin{tabular}{rcc}
\toprule
Class&First binary setting&Second binary setting\\
\midrule
139&$y=2:\{2,3\}$&$y=3:\{1,3\}$\\
166&$y=2:\{2,3\}$&$y=3:\{2,3\}$\\
577&$y=2:\{1,3\}$&$y=3:\{1,3\}$\\
240&$y=1:\{1,3\}$&$y=3:\{2,3\}$\\
242&$y=1:\{1,3\}$&$y=3:\{1,3\}$\\
534&$y=1:\{2,3\}$&$y=3:\{1,3\}$\\
535&$y=1:\{2,3\}$&$y=3:\{2,3\}$\\
\bottomrule
\end{tabular}
\end{center}
The retained PVM score upper bounds for classes 240, 242, 534, and 535 are approximately $9.905100483$, $9.905100469$, $9.905100635$, and $9.905100682$, respectively; their feasible lower scores are approximately $9.905100325$. The second reduction below transfers the full stored PVM upper bounds, with their sampled-span qualifications, to POVMs. No printed rounded endpoint is treated as an exact certificate.
\subsection{An analytical exclusion bound for 139, 166 and 577}
After merging duplicate columns, use class 577 as the common effective witness. The first setting is ternary, with effects $E_1,E_2,E_3$; the other two are binary. Preparations can be taken pure because the score is affine in each preparation separately. Set
\begin{equation}
 R=\rho_1-\rho_4,\qquad X=E_2-E_1,\qquad K=E_3.
\end{equation}
Subtracting the common traceless operator $R$ from the first setting's three reward operators gives $(-R,R,\rho_2+\rho_3)$. The other rewards are
\begin{equation}
 H_2=\rho_1+\rho_2-\rho_3+\rho_4,\qquad
 H_3=\rho_1-\rho_2+\rho_3+\rho_4.
\end{equation}
Each $H_j$ has at most one strictly negative eigenvalue: it is a positive operator minus a pure state. Hence the rewarded binary effects may be chosen $I-P$ and $I-Q$, where $P,Q$ are rank-one or zero projectors. Eigenvalue-zero subspaces can be assigned to the rewarded effect without changing its score. Also,
\begin{equation}\label{reduction:constraints}
 0\preceq K\preceq I,\qquad -(I-K)\preceq X\preceq I-K.
\end{equation}
With $U=P+Q$ and $D=P-Q$, elimination of the four preparations yields the upper bound
\begin{equation}\label{reduction:scorebound}
 S\le4+f(X,-U)+f(D,K),\qquad
 f(H,L)=\lambda_{\max}(L+H)+\lambda_{\max}(L-H).
\end{equation}
Indeed, before eliminating them the four operators paired with $\rho_1,\rho_4,\rho_2,\rho_3$ are respectively $X-U,-X-U,K-D,K+D$, in addition to the constant four. Enlarging the set of preparations at this stage is valid for an upper bound.

\begin{lemma}\label{reduction:spectral}
For Hermitian $H,L$ in dimension at least two, let $\ell_1\ge\ell_2$ be the largest two eigenvalues of $L$, and let $\delta(H)$ be the spectral diameter of $H$. Then
\begin{equation}
 f(H,L)\le\ell_1+\ell_2+
 \sqrt{(\ell_1-\ell_2)^2+\delta(H)^2}.
\end{equation}
\end{lemma}
\begin{proof}
Let $u,v$ be maximizing unit vectors for $L+H,L-H$, and put $r=|\langle u,v\rangle|$. The nonzero eigenvalues of $P_u+P_v$ are $1\pm r$, while those of $P_u-P_v$ are $\pm\sqrt{1-r^2}$. The trace rearrangement inequality bounds the two contributions by
\begin{equation}
 \Tr L(P_u+P_v)\le\ell_1+\ell_2+r(\ell_1-\ell_2),\qquad
 \Tr H(P_u-P_v)\le\delta(H)\sqrt{1-r^2}.
\end{equation}
Cauchy--Schwarz proves the claim. No positivity assumption on $L$ is needed.
\end{proof}

Suppose first that both $P,Q$ have rank one, and put $t=|\langle p,q\rangle|\in[0,1]$. The eigenvalues of $-U$ are $0,-(1-t),-(1+t)$, and $\delta(D)=2\sqrt{1-t^2}$. Let $1\ge k_1\ge k_2\ge k_3\ge0$ be the eigenvalues of $K$. Applying \eqref{reduction:constraints} to orthogonal extremal eigenvectors of $X$ gives
\begin{equation}
 \delta(X)\le2-k_2-k_3.
\end{equation}
Apply Lemma~\ref{reduction:spectral} in \eqref{reduction:scorebound}, and relax $k_1$ upward to one and $k_3$ downward to zero. With $b=k_2$ this gives
\begin{equation}\label{reduction:F}
 S\le F(t,b)=4+t+b+\sqrt{(1-t)^2+(2-b)^2}
                    +\sqrt{(1-b)^2+4(1-t^2)}.
\end{equation}
For fixed $t$, this is convex in $b$, so it suffices to bound the endpoints $b=0,1$.

At $b=0$, set
\begin{equation}
 f_0(t)=4+t+\sqrt{(1-t)^2+4}+\sqrt{5-4t^2}.
\end{equation}
It is concave on $[0,1]$, since
\begin{equation}
 f_0''(t)=\frac{4}{((1-t)^2+4)^{3/2}}-
          \frac{20}{(5-4t^2)^{3/2}}
 \le\frac12-\frac4{\sqrt5}<0.
\end{equation}
Its tangent at $t=2/5$ has negative slope $1-11/\sqrt{109}$. Therefore, for $t\ge0$,
\begin{equation}
 f_0(t)\le f_0(2/5)-\frac25f_0'(2/5)
 =4+\frac{48}{\sqrt{109}}<\frac{43}{5}.
\end{equation}
The final strict inequality follows exactly from $240^2<23^2\cdot109$.

At $b=1$, convexity puts the first square root below its endpoint chord:
\begin{equation}
 \sqrt{(1-t)^2+1}\le\sqrt2+(1-\sqrt2)t.
\end{equation}
Consequently,
\begin{equation}
 F(t,1)\le5+\sqrt2+(2-\sqrt2)t+2\sqrt{1-t^2}
 \le5+\sqrt2+\sqrt{10-4\sqrt2}<\frac{17}{2}.
\end{equation}
For the last inequality, square the positive comparison
$\sqrt{10-4\sqrt2}<7/2-\sqrt2$ and use $17/4>3\sqrt2$.

If precisely one of $P,Q$ is zero, the top two eigenvalues of $-U$ are zero and $\delta(D)=1$. The same bounds give
$S\le7+\sqrt{(1-b)^2+1}\le7+\sqrt2<43/5$.
If both are zero, $f(X,0)=\delta(X)\le2$ and $f(0,K)=2k_1\le2$, giving $S\le8$.
Thus all cases imply
\begin{equation}
 Q_3^{\rm POVM}\le\frac{43}{5},\qquad
 Q_3^{\rm PVM}\le\frac{43}{5},\qquad \eta_*\ge\frac{10}{13}.
\end{equation}
This bound suffices to exclude these classes from improving the reported detection threshold; we do not claim that $43/5$ is attainable.

\subsection{Projective sufficiency for 240, 242, 534 and 535}
After merging duplicate output columns, use class 242. Settings one and three are binary. In setting two the reward operators are
\begin{equation}
 0,\quad 2(\rho_1-\rho_3),\quad \rho_1+\rho_2-\rho_3.
\end{equation}
Subtracting their common traceless shift $R=\rho_1-\rho_3$ gives $(-R,R,\rho_2)$. In particular, $\rho_4$ does not occur in this setting.

\begin{theorem}\label{reduction:pvm}
For any pure qutrit preparations $\rho_1,\rho_2,\rho_3$, the ternary discrimination functional with rewards $(-R,R,\rho_2)$, $R=\rho_1-\rho_3$, has a qutrit projective maximizer. It follows that the complete score optima of classes 240, 242, 534 and 535 satisfy $Q_3^{\rm POVM}=Q_3^{\rm PVM}$.
\end{theorem}
\begin{proof}
On the span of the three preparation vectors, the measurement SDP and its dual have the usual forms
\begin{equation}
 \max_{E_b\succeq0,\,\sum_bE_b=I}\sum_b\Tr(H_bE_b),
 \qquad
 \min_\Gamma\Tr\Gamma\quad\text{subject to }\Gamma\succeq H_b,
 \quad(H_1,H_2,H_3)=(-R,R,\rho_2).
\end{equation}
Both have strict feasible points, so optima exist and complementary slackness applies.

First suppose that the three vectors are linearly independent. Since $\Gamma\succeq\rho_2\succeq0$, $\Gamma$ is positive semidefinite. If $v\in\ker\Gamma$, the constraints $\Gamma\pm R\succeq0$ imply $Rv=0$, and $\Gamma\succeq\rho_2$ implies $\rho_2v=0$. For distinct pure states, $\ker R$ is the orthogonal complement of their two-dimensional span. Independence then forces $v=0$, so $\Gamma\succ0$.

Each $H_b$ has at most one positive eigenvalue. Hence $\dim\ker(\Gamma-H_b)\le1$: on any such kernel, the quadratic form of $H_b$ equals the strictly positive form of $\Gamma$, whereas $H_b$ cannot be strictly positive on a two-dimensional subspace. Complementary slackness gives
\begin{equation}
 \operatorname{supp} E_b\subseteq\ker(\Gamma-H_b),\qquad \operatorname{rank}E_b\le1.
\end{equation}
Three rank-at-most-one positive effects summing to $I_3$ must be three mutually orthogonal rank-one projectors. To see this, write $E_b=w_bw_b^\dagger$; the square matrix with columns $w_b$ satisfies $WW^\dagger=I_3$ and is unitary.

If the preparation span has dimension two and $R\ne0$, then $R$ is invertible on that span. The same kernel argument makes the restricted dual positive definite, so there is an optimal compressed POVM with all three effects of rank at most one. Write its effects $w_bw_b^\dagger$ and let $W$ be the $2\times3$ matrix of columns $w_b$. The relation $WW^\dagger=I_2$ allows its two orthonormal rows to be completed to a $3\times3$ unitary. The three projectors onto that unitary's columns compress to the original effects. This dilation fits inside the available qutrit space and preserves all three reward expectations. If $R=0$, only the reward $\rho_2$ remains, and a projector onto its state, completed to a PVM, is optimal. This also covers a one-dimensional span.

Finally choose pure preparations at a global score optimum. Each binary setting has a projective maximizer by taking the positive spectral subspace of its reward difference. The preceding argument supplies a projective maximizer for the ternary setting without changing the preparations or decreasing the score. The reverse inequality follows from PVMs being a subset of POVMs.
\end{proof}

The theorem concerns the noiseless score. It does not establish equality for direct visibility optimization or for the exact-maximizer visibility convention. Only the PVM score upper bound and its consequence for $\eta=(C_3-6)/(Q_3-6)$ can be transferred; the proof does not transfer PVM visibility bounds to POVMs.

\end{document}